\documentclass[lettersize,journal]{IEEEtran}
\usepackage{amsmath,amsfonts}
\usepackage{algorithmic}
\usepackage{algorithm}
\usepackage{array}
\usepackage[caption=false,font=normalsize,labelfont=sf,textfont=sf]{subfig}
\usepackage{textcomp}
\usepackage{xcolor}
\usepackage{stfloats}
\usepackage{url}
\usepackage{verbatim}
\usepackage{graphicx}
\usepackage{multirow}
\usepackage{pifont}% http://ctan.org/pkg/pifont
\newcommand{\cmark}{\ding{51}}%
\newcommand{\xmark}{\ding{55}}%
\usepackage{booktabs}
\usepackage{cite, hyperref}
\begin{document}

\title{Joint Learning using Mixture-of-Expert-Based Representation for Speech Enhancement and Robust Emotion Recognition}

\author{Jing-Tong Tzeng,~\IEEEmembership{Student Member, IEEE}, Carlos Busso,~\IEEEmembership{Fellow, IEEE}, Chi-Chun Lee,~\IEEEmembership{Senior Member, IEEE}
        % <-this % stops a space

\thanks{Jing-Tong Tzeng and Chi-Chun Lee are with the Department of Electrical Engineering, National Tsing Hua University, Hsinchu 30013, Taiwan (e-mail: roger37890426@gmail.com; cclee@ee.nthu.edu.tw).}% <-this % stops a space
\thanks{Carlos Busso is with the Language Technologies Institute, Carnegie Mellon University, Pittsburgh, PA 15213, USA (e-mail: busso@cmu.edu)}}

% The paper headers
\markboth{Journal of \LaTeX\ Class Files,~Vol.~14, No.~8, August~2021}%
{Shell \MakeLowercase{\emph{et al.}}: A Sample Article Using IEEEtran.cls for IEEE Journals}

\IEEEpubid{0000--0000/00\$00.00~\copyright~2021 IEEE}
% Remember, if you use this, you must call \IEEEpubidadjcol in the second
% column for its text to clear the IEEEpubid mark.

\maketitle

\begin{abstract}
%% The abstract must be between 150-250 words.
\emph{Speech emotion recognition} (SER) plays a critical role in building emotion-aware speech systems, but its performance degrades significantly under noisy conditions. Although \emph{speech enhancement} (SE) can improve robustness, it often introduces artifacts that obscure emotional cues and adds computational overhead to the pipeline. \emph{Multi-task learning} (MTL) offers an alternative by jointly optimizing SE and SER tasks. However, conventional shared-backbone models frequently suffer from gradient interference and representational conflicts between tasks. To address these challenges, we propose the Sparse \emph{Mixture-of-Experts Representation Integration Technique} (Sparse MERIT), a flexible MTL framework that applies frame-wise expert routing over self-supervised speech representations. Sparse MERIT incorporates task-specific gating networks that dynamically select from a shared pool of experts for each frame, enabling parameter-efficient and task-adaptive representation learning. Experiments on the MSP-Podcast corpus show that Sparse MERIT consistently outperforms baseline models on both SER and SE tasks. Under the most challenging condition of -5 dB \emph{signal-to-noise ratio} (SNR), Sparse MERIT improves SER F1-macro by an average of 12.0\% over a baseline relying on a SE pre-processing strategy, and by 3.4\% over a naive MTL baseline, with statistical significance on unseen noise conditions. For SE, Sparse MERIT improves \emph{segmental SNR} (SSNR) by 28.2\% over the SE pre-processing baseline and by 20.0\% over the naive MTL baseline. These results demonstrate that Sparse MERIT provides robust and generalizable performance for both emotion recognition and enhancement tasks in noisy environments.
\end{abstract}


\section{Introduction}
%% Applications of SER -> deploy to real-world scenario  -> noise decreases the SER performance -> noise robustness SER -> not providing enhanced recordings -> trust? -> enhancement than recognition -> not direct optimize -> direct co-train is unstable -> Our proposed method

\IEEEPARstart{S}{peech} emotion recognition (SER) plays a vital role in advancing \emph{Human-Computer Interaction} (HCI) by enabling machines to perceive and respond to human emotions through vocal cues. This capability supports a range of emotion-aware applications, including virtual assistants \cite{zadeh_2023_1, chatterjee_2021_1, chiu_2020_1, kumar_2023_1}, mental health monitoring systems \cite{elsayed_2022_1, madanian_2022_1, harati_2018_1}, and customer service automation platforms \cite{Li_2021_1, han_2020_1, feng_2023_1}. However, speech is often corrupted by background noise in real-world deployment scenarios. Such non-stationary background noise can obscure emotion-relevant acoustic features and significantly degrade SER performance, thereby limiting its reliability and generalizability.

To improve the noise robustness of SER, numerous approaches have been explored, including robust feature engineering \cite{chakraborty_2019_1, schuller_2006_1, leem_2022_1}, data augmentation \cite{jaiswal_2021_1, ranjan_2024_1, tiwari_2020_1}, environment-dependent compensations \cite{Leem_2023_2, Leem_202x}, and domain adaptation \cite{wilf_2021_1, leem_2021_1, leem_2023_1}. While these strategies have demonstrated effectiveness in enhancing SER performance under noisy conditions, their inability to produce cleaned speech limits their usefulness in applications that require human intervention or audio auditing, such as emergency response systems \cite{deschamps_2021_1, deb_2023_1, deschamps_2022_1}. In such scenarios, access to intelligible speech is as critical as accurate emotion recognition. For example, human operators may need to directly review the spoken content to make informed decisions, assess urgency, or validate automated predictions. Given these limitations, \emph{speech enhancement} (SE) offers a more interpretable and versatile solution by generating denoised speech that supports both automated processing and human-in-the-loop analysis \cite{triantafyllopoulos_2019_1, kshirsagar_2023_1, chen_2023_1}. However, SE models are typically optimized for perceptual intelligibility and signal fidelity, objectives that do not necessarily align with preserving the emotion-discriminative features needed for SER \cite{avila_2018_1}. As a result, emotional nuances may be unintentionally suppressed during enhancement. Additionally, incorporating SE as a standalone front-end module increases model complexity and computational overhead, which can hinder its practical deployment in resource-constrained environments.

In our previous work \cite{tzeng_2025_1}, we addressed this mismatch and computational overhead by jointly training SE and SER models using shared self-supervised speech pre-trained model representations. This \emph{multi-task learning} (MTL) framework improved noise robustness while reducing model redundancy. However, MTL models with a single shared backbone often suffer from unstable training dynamics. As noted in prior studies \cite{chen_2018_1, yu_2020_1, ma_2018_1}, shared parameters can receive conflicting gradient signals from different task objectives, leading to suboptimal convergence and biased feature representations. This issue is further compounded when the tasks differ significantly in complexity. For example, speech enhancement requires fine-grained, low-level signal reconstruction, whereas speech emotion recognition involves high-level abstraction and semantic understanding. These differences pose a challenge for a single backbone to serve both tasks effectively, often resulting in suboptimal convergence and degraded performance.

\IEEEpubidadjcol

This paper proposes the Sparse \emph{Mixture-of-Experts Representation Integration Technique} (Sparse MERIT), a flexible MTL framework designed to integrate speech self-supervised representations for both SE and SER. Sparse MERIT addresses the limitations of conventional shared-backbone architectures by incorporating a \emph{Mixture-of-Experts} (MoE) structure that expands model capacity and enables more effective representation integration across tasks. Rather than relying on a single shared pathway, Sparse MERIT introduces multiple expert modules along with task-specific gating networks that dynamically select expert outputs based on the input. This design mitigates negative interference between conflicting objectives and better accommodates the different levels of complexity required by SE and SER tasks.

Experiments on the MSP-Podcast corpus \cite{lotfian_2017_1} demonstrate that Sparse MERIT improves generalization for both tasks, particularly under challenging noisy conditions. Under the most difficult setting of -5 dB \emph{signal-to-noise ratio} (SNR), Sparse MERIT outperforms baseline relying on a SE pre-processing strategy by an average of 12.0\% F1-macro. It also improves upon our previously proposed naive MTL framework by 3.4\%, across two unseen noise datasets, with statistical significance. In addition, Sparse MERIT consistently improves SE performance across multiple standard enhancement metrics. These results confirm that jointly learning SE and SER through our Sparse MERIT architecture leads to more robust and effective performance than prior MTL strategies.

The main contributions of this paper are summarized as follows:
\begin{itemize}
\item We show that combining SE and SER in a multi-task framework improves both enhancement quality and emotion recognition performance under diverse noise conditions.
\item We introduce Sparse MERIT, a flexible MoE-based architecture that goes beyond our prior work by enhancing representation capacity and reducing task interference via task-specific expert routing.
\item We validate Sparse MERIT through extensive experiments, showing consistent gains over a SE pre-processing baseline, a naive MTL baseline, and other mainstream techniques, especially in unseen noise conditions.
\end{itemize}

The rest of this paper is organized as follows. Section \ref{sec:related_work} reviews related work on SER in noisy conditions and MTL strategies. Section \ref{sec:proposed_method} introduces the proposed Sparse MERIT framework, including its representation integration, expert routing mechanism, and task-specific components. Section \ref{sec:experimental_settings} describes the experimental setup, including datasets, implementation details, and baseline comparisons. Section \ref{sec:results} presents the results and analysis for both SER and SE tasks, along with ablation studies. Finally, Section \ref{sec:conclusion} concludes the paper and discusses directions for future work.

\section{Related Works} 
\label{sec:related_work}
\subsection{Speech Emotion Recognition under Noisy Conditions}
%%feature selection->frame selection->data augmentation->domain adaption->speech enhancement

Recent studies have demonstrated significant progress in SER \cite{Goncalves_2024}. However, SER systems remain highly vulnerable in noisy environments, posing a major barrier to their deployment in real-world applications. One line of work focuses on noise-robust feature selection. For example, Schuller \emph{et al.} \cite{schuller_2006_1} applied information gain ratio-based feature selection and demonstrated improved performance under both clean and noisy conditions. Leem \emph{et al.} \cite{leem_2022_1} identified a subset of noise-robust \emph{low-level descriptors} (LLDs), which outperformed the full LLD set in noisy settings. Building on this idea, Leem \emph{et al.} \cite{leem_2024_1} proposed a \emph{generative adversarial network} (GAN)-based feature enhancement model that strengthens weak features while preserving robust ones. Similarly, Chakraborty \emph{et al.} \cite{chakraborty_2019_1} employed a denoising autoencoder to enhance \emph{Mel-Frequency Cepstral Coefficient} (MFCC) features, achieving improvements in robustness. 

Another direction improves SER by discarding noisy frames. Pandharipande \emph{et al.} \cite{pandharipande_2018_1, pandharipande_2018_2} used a front-end \emph{voice activity detector} (VAD) to identify and discard noisy frames prior to feature extraction. Leem \emph{et al.} ~\cite{leem_2024_2} extended this approach by replacing dropped frames with enhanced speech, thereby preserving lexical content and improving recognition accuracy.

A third strategy focuses on increasing data diversity by contaminating clean training speech with various noise types. This approach exposes the model to a wider range of acoustic conditions during training. Tiwari \emph{et al.} \cite{tiwari_2020_1} proposed a generative model capable of synthesizing diverse noise profiles in the Mel-filterbank energy domain. Ranjan \emph{et al.} \cite{ranjan_2024_1} developed a \emph{reinforcement learning} (RL)-based augmentation method that adaptively chooses noise types to optimize performance under unseen conditions.

A fourth line of work incorporates environmental information directly into the model to enhance noise robustness. Leem \emph{et al.} \cite{Leem_2023_2} proposed skip-connection adapters composed of environment-agnostic and environment-specific modules to denoise speech representations within a transformer encoder. Additionally, they used text-based environment descriptions to further enrich the contextual representation and improve robustness in their later work \cite{Leem_202x}.

Another direction treats noise robustness as a domain mismatch problem. Leem \emph{et al.} \cite{leem_2021_1} employed a ladder network \cite{valpola_2015_1, parthasarathy_2020_1}, in which the final-layer embeddings are separated into two branches: one for emotion classification and the other for reconstructing clean speech representations. This dual-branch design promotes discriminative features while mitigating background noise. In a subsequent study, Leem \emph{et al.} \cite{leem_2023_1} proposed a contrastive teacher–student framework to align noisy embeddings with their clean counterparts, thereby improving generalization to unseen noise conditions. Liu \emph{et al.} \cite{liu_2025_1} leveraged a diffusion denoising probabilistic model to transfer emotional information from clean to noisy speech with an iterative strategy that progressively adapts the classification model to noisy feature distributions. Wilf and Provost \cite{wilf_2021_1} introduced a MoE architecture integrated with a \emph{Domain Separation Network} (DSN) \cite{Bousmalis_2016_1}, enabling input-dependent routing to specialized encoders based on noise characteristics and enhancing robustness in both unimodal and multimodal settings.

However, the aforementioned approaches do not generate enhanced speech signals that can be inspected by humans, which limits their practical utility in real-world settings. Using a front-end SE module has been explored as a more practical solution, as it not only improves SER performance but also increases transparency and user trust by providing human-interpretable, denoised speech signals. Triantafyllopoulos \emph{et al.} \cite{triantafyllopoulos_2019_1} incorporated SE as a front-end component to improve SER performance, particularly under low SNR conditions. Kshirsagar \emph{et al.} \cite{kshirsagar_2023_1} employed front-end SE with a mimic loss \cite{bagchi_2018_1} originally developed for \emph{automatic speech recognition} (ASR), and demonstrated improved SER performance in a multimodal framework. Chen \emph{et al.} \cite{chen_2023_1} proposed an SNR-level detection module to reduce the aliasing effects of SE on speech signals with little or no background noise. To further explore the interaction between SE and SER, Avila \emph{et al.} \cite{avila_2018_1} investigated the correlation between perceptual speech quality and emotion classification accuracy. Despite their effectiveness, these two-stage approaches are often resource-intensive, increasing model complexity and limiting their suitability for deployment in resource-constrained settings. Moreover, the perceptual speech quality metrics used in the first stage are not specifically designed to capture emotional cues, which may result in a mismatch between enhancement objectives and the needs of emotion recognition.

\subsection{Multi-Task Learning}
 %%uncertainty loss -> gradient surgery -> mixture of expert
To reduce computational cost and address the mismatch between speech intelligibility and model recognition performance, MTL offers a promising solution by enabling the joint optimization of multiple objectives. Several studies have shown that incorporating reconstruction loss as an auxiliary objective can enhance SER performance \cite{wilf_2021_1, khorram_2017_1, leem_2021_1, goncalves_2025_1}. However, these approaches primarily focus on reconstructing intermediate feature representations rather than the waveform itself. As a result, their practical utility in applications requiring human-audible outputs remains limited. 

Speech \emph{self-supervised learning} (SSL) models have shown strong performance across speech tasks \cite{baevski_2020_1, hsu_2021_1, chen_2022_1, yang_2021_1, tsai_2022_1}, including both SE \cite{hung_2022_1, song_2023_1, zhao_2022_1} and SER \cite{wang_2021_1, atmaja_2022_1, naini_2025_1}. Therefore, there is a strong motivation to adopt a unified SSL backbone for MTL involving both tasks. Our previous work further supports this direction, showing that jointly learning SE and SER from shared SSL representations improves SER robustness under unseen noisy conditions without compromising SE performance \cite{tzeng_2025_1}. 

Although MTL offers potential benefits through shared representation learning, MTL models do not always outperform their single-task counterparts across all tasks in practice \cite{luong_2015_1, kaiser_2017_1}. This inconsistency is often attributed to several inherent challenges, including gradient interference between tasks, training instability, and imbalanced learning dynamics caused by differences in task complexity. To address these issues, various strategies have been proposed. One such approach is uncertainty-based loss weighting \cite{kendall_2018_1}, which introduces task-dependent homoscedastic uncertainty as learnable parameters to dynamically adjust the contribution of each task’s loss. This strategy allows the model to adaptively balance the competing objectives, without requiring manual loss reweighting.

Another line of work focuses on directly manipulating task gradients to address training instability and reduce conflicts in multi-task optimization. GradNorm \cite{chen_2018_1} is a representative example that balances learning across tasks by dynamically adjusting gradient magnitudes based on the relative training speed of each task. By equalizing the rate at which task-specific losses decrease, GradNorm helps prevent any single task from dominating the optimization process. In contrast, \emph{projected conflicting gradient} (PCGrad) \cite{yu_2020_1} addresses gradient interference by projecting out the conflicting components between task gradients, reducing destructive updates and improving training stability.

Compared to approaches that automatically balance task losses or adjust gradient magnitudes and directions to stabilize training, MoE architectures offer an alternative solution to MTL through architectural design \cite{ma_2018_1, gupta_2022_1}. An MoE framework typically consists of a shared pool of expert networks and a gating mechanism that determines which subset of experts to activate for a given input. This conditional routing mechanism increases model capacity without proportional computational overhead and allows the model to learn more flexible, input- or task-sensitive processing paths. Early applications of MoE in MTL used sample-level routing, where each input is assigned to a subset of experts. For example, Wilf and Provost \cite{wilf_2021_1} applied an MoE model with a noise-type classifier to dynamically route inputs to different feature encoders, performing both SER and feature reconstruction to improve noise robustness. While effective, sample-level routing often lacks granularity because it assumes a uniform expert assignment across all frames of an utterance. This strategy can lead to suboptimal performance when local acoustic or emotional variations are present. To address this limitation, token-level MoE has emerged as a more flexible alternative, allowing each token to be routed independently based on its local representation. In the domain of language modeling, approaches such as the Switch Transformer \cite{fedus_2022_1} have shown that sparse token-wise MoE architectures can significantly scale model capacity without increasing inference cost. This design has been widely adopted in \emph{large language models} (LLMs) \cite{jiang_2024_1, yang_2025_1, zhu_2024_1, dai_2024_1}, where it improves both computational efficiency and model expressiveness. These studies highlight the potential benefits of sparse token-level MoE, motivating its adoption for speech-based multi-task learning.

Token-wise MoE has also been adapted to MTL settings, where it helps support task heterogeneity and feature specialization. For instance, Liang \emph{et al.} \cite{liang_2022_1} introduced M$^3$ViT for vision tasks and showed that sparse patch-level expert selection improves multi-task performance. In the speech domain, frame-wise MoE has also been effective. You \emph{et al.} \cite{you_2021_1, you_2022_1} applied MoE to speech recognition with favorable results. Further improvements have been demonstrated in multilingual speech recognition \cite{wang_2023_1, hu_2023_1}.

Building on these advances, we propose Sparse MERIT, an MoE-based framework designed for MTL over speech self-supervised representations, targeting both SER and SE. Sparse MERIT leverages dynamic expert routing at the frame level to reduce gradient interference, support parameter-efficient specialization across tasks, and improve generalization without increasing inference cost.

\section{Proposed Method}
\label{sec:proposed_method}

\begin{figure*}[!t]
\includegraphics[width=0.85\textwidth]{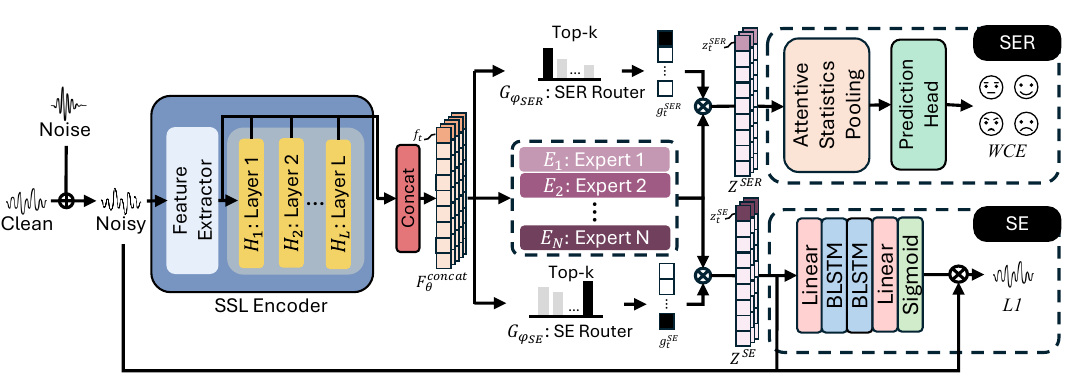}
\centering
\caption{The proposed Sparse MERIT framework for enhanced speech emotion recognition, leveraging unified self-supervised speech representations through token-wise expert routing.}
\label{fig:framework}
\end{figure*}

This section outlines the architecture of Sparse MERIT, our proposed MTL framework for SER and SE. Sparse MERIT builds on our preliminary work \cite{tzeng_2025_1} by introducing a frame-wise MoE layer over multi-layer self-supervised speech representations. This framework consists of three main components: (1) a layer-wise feature construction from a pre-trained SSL model, (2) an expert-based integration using frame-level sparse routing, and (3) task-specific heads for SE and SER trained under a joint objective. Figure \ref{fig:framework} illustrates the overall architecture.

\vspace{-4pt}
\subsection{Layer-Wise Representation Construction}

Given a noisy input waveform \( x_{\text{noisy}} \in \mathbb{X}_{\text{noisy}} \) and its corresponding clean reference waveform \( x_{\text{clean}} \in \mathbb{X}_{\text{clean}} \), we extract hidden representations from a pre-trained SSL model parameterized by \( \theta \). The SSL model consists of an input feature extractor followed by \( L \) transformer layers. Let \( H_0 \in \mathbb{R}^{T \times D} \) denote the input to the first transformer layer, and let \( S_{\theta}^{l}(x_{\text{noisy}}) \in \mathbb{R}^{T \times D} \) denote the output of the \( l \)-th transformer layer for \( l = 1, \ldots, L \), where \( T \) is the number of frames and \( D \) is the feature dimensionality.

For notational convenience, we define:
\begin{equation}
H_l = S_{\theta}^{l}(x_{\text{noisy}}), \quad \text{for } l = 1, \ldots, L
\end{equation}

We construct a comprehensive multi-layer representation by concatenating the input \( H_0 \) and all transformer outputs \( H_1, \ldots, H_L \) along the feature dimension:
\begin{equation}
F^{\text{concat}}_{\theta} = \text{Concat}(H_0, H_1, H_2, \ldots, H_L) \in \mathbb{R}^{T \times ((L+1) \cdot D)}
\end{equation}

% To construct a comprehensive multi-layer representation, we use the outputs from all \( L \) layers, since each layer captures different levels of information as discussed in \cite{upadhyay24_interspeech}. In this study, we fuse the feature representations from all layers through concatenation, defined as:
% \begin{equation}
% F^{\text{concat}}_{\theta} = \text{Concat}(S_{\theta}^{0}(x_{\text{noisy}}), \ldots, S_{\theta}^{L}(x_{\text{noisy}})) \in \mathbb{R}^{T \times (L \cdot D)}
% \end{equation}

This frame-level sequence captures multi-scale contextual information across multiple abstraction levels, serving as input to the MoE module.

\subsection{Mixture-of-Experts Integration}

To process the concatenated multi-layer representation \( F_{\theta}^{\text{concat}} \), we introduce a frame-wise MoE module. Each frame embedding \( {f}_t \in \mathbb{R}^{(L+1) \cdot D} \), corresponding to the \( t \)-th frame of \( F_{\theta}^{\text{concat}} \), is routed to one of \( N \) shared expert networks \( \{E_n\}_{n=1}^{N} \).

Each expert \( E_n \) is implemented as a two-layer feedforward network that projects the high-dimensional input into a lower-dimensional embedding space of size \( D \). This mapping reduces the concatenated feature dimensionality while preserving temporal resolution:
\begin{equation}
E_n: \mathbb{R}^{(L+1) \cdot D} \rightarrow \mathbb{R}^{D}
\end{equation}

To enable task-specific routing, we introduce two independent gating networks, \( G_{\phi_{\text{SER}}} \) and \( G_{\phi_{\text{SE}}} \), parameterized by \( \phi_{\text{SER}} \) and \( \phi_{\text{SE}} \), respectively. For each frame \( t \), the gating network for task \( {\tau} \in \{\text{SER}, \text{SE}\} \) produces a softmax-normalized routing score over the \( N \) experts:
\begin{equation}
{g}_t^{\tau} = \text{softmax}(G_{\phi_{\text{task}}}({f}_t)) \in \mathbb{R}^{N}
\end{equation}

We use a Top-\( K \) routing strategy to select the most relevant experts for each frame. The gating network outputs a probability distribution over experts, and the TopK operator selects the top-\( K \) values, zeroing out the rest. This approach allows each frame to be processed by a sparse subset of experts, which improves computational efficiency and encourages expert specialization. While Sparse MERIT supports arbitrary \( K \geq 1 \), we adopt \( K = 1 \) in this study, following the sparse routing design of the Switch Transformer~\cite{fedus_2022_1}.

Formally, let \( \text{TopK}(v, k) \in \mathbb{R}^N \) denote the operator that retains the top \( k \) values of a vector \( v \), setting the remaining entries to zero:
\begin{equation}
\text{TopK}(v, k)_n = 
\begin{cases}
v_n, & \text{if } v_n \text{ is among the top-}k \text{ elements of } v \\
0, & \text{otherwise}
\end{cases}
\end{equation}

Using Top-\( K \) routing, the MoE output for frame \( t \) and task $\tau$ is computed as a weighted combination of expert outputs, where each expert's output is scaled by its corresponding gating weight:
\begin{equation}
\mathbf{z}_t^{{\tau}} = \sum_{n=1}^{N} \text{TopK}(\mathbf{g}_t^{{\tau}}, K)_n \cdot {E}_n({f}_t)
\label{eq:topKeps}
\end{equation}

The final MoE output sequence for task \(\tau\) is thus:
\begin{equation}
Z^{{\tau}} = [\mathbf{z}_1^{{\tau}}, \mathbf{z}_2^{{\tau}}, \ldots, \mathbf{z}_T^{{\tau}}] \in \mathbb{R}^{T \times D}
\end{equation}

This design enables task-specific expert selection at the frame level, balancing specialization and parameter sharing, while ensuring a consistent output shape across both tasks. 

\subsection{Task-Specific Heads}

\subsubsection{Speech Emotion Recognition}
The SER task takes the MoE-transformed sequence $Z^{\text{SER}} \in \mathbb{R}^{T \times D}$ as input. We apply attentive statistics pooling \cite{okabe_2018_1} to convert frame-level features into a fixed-length utterance representation. This strategy is a temporal aggregation method that computes the weighted mean and standard deviation across time using learned attention weights. This pooled vector is passed to a task-specific classification head $\Pi_{\text{SER}}$, parameterized by $\theta_{\text{SER}}$, to predict the emotionl label:

\begin{equation}
\hat{y} = \Pi_{\text{SER}}(\text{Pooling}(Z^{\text{SER}})), \quad \hat{y} \in \mathbb{Y}
\end{equation}

\subsubsection{Speech Enhancement}
For the SE task, we first compute a spectral representation from the noisy waveform using the magnitude of the \emph{short-time Fourier transform} (STFT). The result is then compressed using the log1p function, defined as $log1p(x)=\log(1 + x)$, which has been shown to improve SE performance \cite{fu_2020_1}:

\begin{equation}
X_{\text{noisy}} = \log(1 + |\text{STFT}(x_{\text{noisy}})|)
\end{equation}

We concatenate this spectral feature with the MoE-transformed output $Z^{\text{SE}}$ along the feature dimension and feed it to the SE head $\Pi_{\text{SE}}$, parameterized by $\theta_{\text{SE}}$, to reconstruct the enhanced spectrogram:

\begin{equation}
\hat{X}_{\text{clean}} = \Pi_{\text{SE}}(Z^{\text{SE}}, X_{\text{noisy}})
\end{equation}

\subsection{Multi-Task Objective}

We jointly optimize the SE and SER tasks using an MTL objective. The model is trained to minimize the sum of a weighted cross-entropy loss for SER and an $L_1$ loss for SE:

\begin{equation}
\mathcal{L}=\min_{\theta, \phi_{\text{SER}}, \phi_{\text{SE}}, \theta_{\text{SER}}, \theta_{\text{SE}}} \  \mathcal{L}_{\text{WCE}}(\hat{y}, y) + \mathcal{L}_1(\hat{X}_{\text{clean}}, X_{\text{clean}})
\end{equation}

The weighted cross-entropy loss $\mathcal{L}_{\text{WCE}}$ is used to compensate for the imbalanced class distribution in the emotion dataset, ensuring that underrepresented categories are not neglected during training. The $\mathcal{L}_1$ loss encourages accurate reconstruction of clean spectral features for speech enhancement. This combined objective guides the model to learn representations that support both high-level semantic discrimination (emotion classification) and low-level signal reconstruction (enhancement), while allowing shared learning through the unified self-supervised backbone.

\section{Experimental Settings}
\label{sec:experimental_settings}
\subsection{Data Preparation}

We conduct our experiments using the MSP-Podcast corpus \cite{lotfian_2017_1}, a large-scale, naturalistic emotional speech dataset derived from a diverse range of podcast recordings. The selected utterances, ranging from 2.75 to 11 seconds in duration, are carefully filtered to exclude background music and overlapping speech. To ensure acoustic quality, only recordings with a predicted SNR above 20 dB are retained. For this study, we focus on four emotion categories: anger, sadness, happiness, and neutral state. We utilize version 1.11 of the corpus, which contains 100,896 labeled segments (Anger: 10,342; Sadness: 8,347; Happiness: 29,454; Neutral: 52,753). The training partition is used to fine-tune a pre-trained speech representation model, and the development set is employed for model selection and early stopping. We evaluate the results on the test 1 set of the corpus. 

To introduce realistic noise conditions during training, we augment the clean data by generating babble noise through speech overlay using samples from the CRSS-4ENGLISH-14 corpus \cite{tao_2018_1}. The training and development sets are corrupted at an SNR of 5 dB to simulate moderate background interference. For evaluation, we apply the same corruption process to the test 1 set using 4 SNR levels: -5 dB, 0 dB, 5 dB, and 10 dB, covering a range of low to high noise intensities. To further test the robustness of the model against unseen noise types, we introduce ambient noise collected from the Freesound repository \cite{fonseca_2017_1}, using the same SNR levels for consistency. Additionally, we incorporate noise samples from the ICASSP 2023 \emph{Deep Noise Suppression} (DNS) Challenge dataset \cite{dubey_2024_1}, which includes diverse real-world noise recordings. To avoid data redundancy, we remove overlapping segments between DNS and Freesound. For experimental simplicity and to isolate noise effects, we exclude room impulse responses from the DNS samples.

\subsection{Implementation Details}

We implement our proposed Sparse MERIT framework\footnote{\scriptsize Code will be publicly available at: \url{https://github.com/RogerTzeng/Sparse-MERIT}} using the WavLM Large model \cite{chen_2022_1} as the shared self-supervised backbone. WavLM Large is a 24-layer transformer model pre-trained on 94K hours of both clean and noisy speech. It has demonstrated strong performance across a wide range of speech processing tasks, including SER and SE, as shown in the \emph{speech processing universal performance benchmark} (SUPERB) \cite{yang_2021_1, tsai_2022_1} and recent works \cite{naini_2025_1, hung_2022_1}. This versatility, along with its robustness to noise, makes WavLM well-suited for our MTL framework.

Following the WavLM backbone, we apply our proposed MoE integration layer to process the concatenated multi-layer representations. The input to the MoE has a dimensionality of $1,024 \times 25$, formed by concatenating the hidden states from the input to the transformer encoder (pre-layer representation) together with the outputs from all 24 transformer layers (each of dimensionality 1,024). The MoE module consists of $N=3$ experts. Each expert first reduces the input dimension from $25 \times 1,024$ to 4,096 and then projects it into a 1,024-dimensional output. This design compresses the high-dimensional concatenated input into a compact task-adapted representation suitable for the downstream processing.

For the SER task, the MoE output is passed to a task-specific classification head composed of attentive statistics pooling \cite{okabe_2018_1} followed by fully connected layers, based on the baseline from the Interspeech 2025 Challenge on Speech Emotion Recognition in Naturalistic Conditions \cite{naini_2025_1}. For the SE task, the MoE output is concatenated with a log-compressed spectrogram of the noisy input. The combined representation is processed by the SE decoder, which adopts the architecture of the BSSE-SE model \cite{hung_2022_1}, designed to reconstruct clean spectral features in noisy environments. 

During preprocessing, all input waveforms are normalized using the Z-normalization, with mean and standard deviation estimated from the entire training set. We use a two-phase training procedure. In the first phase, we freeze the SSL backbone and train the SE and SER heads independently. The SE head is trained using the AdamW optimizer for 130 epochs with a batch size of 16 and a learning rate of $5 \times 10^{-5}$. The SER head is trained for 20 epochs with a batch size of 32 and the same optimizer settings. In the second phase, we jointly fine-tune the full model using the pre-trained head weights. The full pipeline is trained for an additional 20 epochs with a batch size of 32. We continue using the AdamW optimizer, setting the learning rate to $5 \times 10^{-5}$ for the expert networks, gating networks, and task-specific heads. We set the learning rate to $2.5 \times 10^{-5}$ for the Transformer layers of the SSL model. The CNN-based feature extractor of the WavLM backbone remains frozen during both training phases.

\subsection{Baseline Methods for SER}

We compare our proposed method, Sparse MERIT, with nine SER baselines:  
\begin{itemize}
    \item \textbf{Original}: Fine-tunes the SER model on clean emotional speech without any adaptation to noisy conditions.

    \item \textbf{SE Pre-process with BSSE-SE (SE-P w/ BSSE-SE)}: Uses the BSSE-SE model, which is the same SE module adopted in our MTL framework, as a front-end enhancer. BSSE-SE is trained on VCTK-DEMAND \cite{valentini_2016_1} and fine-tuned on MSP-Podcast, and the SER model is subsequently trained using the enhanced speech.
    
    \item \textbf{SE-P w/ SGMSE+}: Replaces the BSSE-SE front-end in SE-P with SGMSE+ \cite{richter_2023_1}, a diffusion-based speech enhancement model, to examine whether a stronger generative enhancer yields additional robustness gains for SER under noisy conditions.
    
    \item \textbf{SE-P w/ SEMamba}: Replaces the BSSE-SE front-end in SE-P with SEMamba \cite{chao_2024_1}, which adopts state-space sequence modules as an alternative to attention-centric designs. This baseline evaluates whether the state-space sequence modeling and its computational characteristics in the SE front-end translate into improved robustness for downstream SER.
    
    \item \textbf{Fine-tuning Entire Model (FT-M)}: Fine-tunes both the SSL backbone and the SER classification head directly on noisy speech data, effectively treating noise corruption as data augmentation to improve robustness.
    
    \item \textbf{Naive Fine-tuning w/ Multi-task Learning (FT-MTL)}: Jointly trains SE and SER using a shared SSL backbone, where a weighted sum of layer-wise representations is used as the input to both task-specific heads. The model is trained on noisy speech by combining the enhancement and classification losses, following the approach proposed in \cite{tzeng_2025_1}.
    
    \item \textbf{FT-MTL w/ Uncertainty}: Extends FT-MTL by applying task uncertainty-based loss weighting \cite{kendall_2018_1} to automatically balance the SE and SER objectives during training.
    
    \item \textbf{FT-MTL w/ PCGrad}: Builds on FT-MTL by applying PCGrad \cite{yu_2020_1} to mitigate gradient interference between tasks and improve training stability.
\end{itemize}
% {\color{red} [the highlighed section suggests that both approaches are our proposed approach. However, section 3 describes sparse MERIT, not Dense MERIT. I Suggest to rephrase this part with the text written below.]

% We also include two variants of our proposed MERIT framework for comparison:

% \begin{itemize}
%     \item \textbf{Dense MERIT}: Implements the MERIT framework with dense expert selection, where each frame-level representation is routed to all experts with continuous soft weights. This allows all experts to contribute to every frame.
    
%     \item \textbf{Sparse MERIT}: Our proposed method, which applies Top-1 frame-wise expert routing. Each frame is processed by a single selected expert, enabling efficient task-adaptive specialization with reduced computational overhead.
% \end{itemize}

% *******  Replace by **********

In addition to our proposed Sparse MERIT approach, we implement a variation to evaluate our decision to only use the Top-1 frame-wise expert routing, where each frame is processed by a single selected expert, enabling efficient task-adaptive specialization with reduced computational overhead. 

\begin{itemize}
    \item \textbf{Dense MERIT}: Implements the MERIT framework with dense expert selection, where each frame-level representation is routed to all experts with continuous soft weights. This allows all experts to contribute to every frame.
\end{itemize}
% }

\subsection{Baseline Methods for SE}
Although SER is the primary task of interest, the quality of the enhanced speech is also crucial for real-world applications in which humans may interact with or listen to the audio output. Poor enhancement quality can degrade recordings' quality and hinder both human understanding and downstream processing. Moreover, evaluating SE performance provides insight into how well an MTL method resolves conflicts between competing objectives. Since SE and SER often require different feature characteristics, joint training can lead to suboptimal performance if the model fails to disentangle the two tasks. Therefore, we compare SE performance across various MTL strategies, as well as models fine-tuned solely for speech enhancement, to assess their effectiveness in mitigating task interference and preserving signal quality.

\begin{itemize}

    % \item \textbf{Noisy}: The noisy speech is contaminated by overlaying speech from the CRSS-4ENGLISH-14 corpus, adding ambient noise from the Freesound repository, or injecting diverse noise types from the DNS Challenge dataset. {\color{red}[how is this a baseline for SE?]}
    \item \textbf{Fine-tuned (SE-only)}: Enhanced speech produced by SE models fine-tuned exclusively for the SE task on MSP-Podcast data contaminated with recordings from the CRSS-4ENGLISH-14 training set. We report results for three SE architectures: BSSE-SE, SGMSE+, and SEMamba. Each model is initialized from a checkpoint pre-trained on VCTK-DEMAND and optimized without any SER objective.

    \item \textbf{FT-MTL Variants}: Speech enhancement outputs generated from jointly trained SE+SER models are considered here. This includes standard FT-MTL and its variants with uncertainty weighting, PCGrad, and both Dense and Sparse MERIT integration strategies.
\end{itemize}

% {\color{red}[not all these models are baselines for SE (e.g., noisy). This subsection reads odd]}

\section{Results}
\label{sec:results}
%% Comparison with the classification baselines, enhanced audio quality, our enhanced recordings preserve more emotion information. How many experts are the best?

\subsection{Noise-Robust Speech SSL Backbone}
Before introducing any explicit robustness-oriented training strategies, we examine the intrinsic noise robustness of different speech SSL backbones. We consider three widely used SSL models for SER: HuBERT \cite{hsu_2021_1, li_2024_1}, wav2vec~2.0 \cite{baevski_2020_1, Leem_2023_2, bijoy_2025_1}, and WavLM \cite{chen_2022_1, mote_2025_1}. To ensure a fair comparison, we fine-tune identical SER classifiers on clean emotional speech and vary only the SSL backbone. We then evaluate these models on the CRSS-contaminated test set to measure how well each backbone generalizes to additive noise without being exposed to noisy training data.

\begin{table}[]
\centering
\caption{SER performance (F1-macro) of models using different SSL backbones evaluated on the CRSS-contaminated test set. The best results are highlighted in bold.}
\begin{tabular}{cccc}
\hline
SNR   & HuBERT & wav2vec~2.0 & WavLM \\ \hline
-5 dB & 0.136 & 0.233 & \textbf{0.287} \\
0 dB  & 0.191 & 0.311 & \textbf{0.410} \\
5 dB  & 0.303 & 0.406 & \textbf{0.506} \\
10 dB & 0.422 & 0.502 & \textbf{0.555} \\ \hline
\end{tabular}
\label{table:ssl_backbone}
\end{table}

As shown in Table~\ref{table:ssl_backbone}, WavLM consistently delivers the highest F1-macro across all SNR conditions, indicating stronger inherent robustness to background noise. A plausible reason is that WavLM's pre-training recipe exposes the model to noisy and overlapping-speech perturbations, which encourages the learned representations to be less sensitive to background interference. Motivated by these results, we use WavLM as the default SSL backbone in the remainder of our experiments.

\subsection{Emotion Recognition}

\begin{table*}[t]
\caption{SER performance of the proposed method (Sparse MERIT) and all baselines. We use symbols to denote when a model performs significantly better than the Original ($\star$), FT-M ($\ast$), SE-P w/ BSSE-SE ($\dag$), SE-P w/ SGMSE+ ($\square$), SE-P w/ SEMamba ($\triangle$), FT-MTL ($\ddag$), FT-MTL w/ Uncertainty ($\circ$), FT-MTL w/ PCGrad ($+$), and Dense MERIT ($\diamond$) models. The best results are highlighted in bold.}

\begin{center}
\resizebox{0.82\linewidth}{!}{
\begin{tabular}{@{}cccccccc@{}}
\toprule
 &              & \multicolumn{2}{c}{CRSS-4ENGLISH-14 (Seen Noise)} & \multicolumn{2}{c}{Freesound (Unseen Noise)} & \multicolumn{2}{c}{DNS (Unseen Noise)} \\ \midrule
SNR                    & Model                 & F1-Macro       & F1-Micro       & F1-Macro       & F1-Micro       & F1-Macro       & F1-Micro       \\ \midrule
\multirow{8}{*}{-5 dB} & Original              & 0.299          & 0.481          & 0.426          & 0.547          & 0.450          & 0.560          \\
                       & FT-M                  & $0.386^{\star}$          & $0.512^{\star}$          & $0.472^{\star}$          & 0.557          & $0.474^{\star}$          & 0.551          \\
                       & SE-P w/ BSSE-SE \cite{hung_2022_1}       & $\textbf{0.507}^{\star\ast}$ & $\textbf{0.579}^{\star\ast}$ & $0.435^{\star}$          & 0.542          & 0.441          & 0.539          \\
                       & SE-P w/ SGMSE+ \cite{richter_2023_1}      & $0.392^{\star}$ & $0.524^{\star\ast}$ & $0.459^{\star\dag}$    &     $0.565^{\star\ast\dag}$   &   0.442        &  0.544         \\
                       & SE-P w/ SEMamba \cite{chao_2024_1}       & $0.437^{\star\ast\square}$ & $0.560^{\star\ast\square}$ &  $0.441^{\star}$         &   $0.558^{\star\dag}$     &  0.433      &   0.543     \\
                       & FT-MTL \cite{tzeng_2025_1}                & $0.388^{\star}$          & $0.512^{\star}$          & $0.471^{\star\dag\square\triangle}$          & $0.573^{\star\ast\dag\square\triangle}$          & $0.478^{\star\dag\square\triangle}$          & $0.569^{\star\ast\dag\square\triangle}$          \\
                       & FT-MTL w/ Uncertainty & $0.390^{\star}$          & $0.527^{\star\ast\ddag}$          & $0.480^{\star\dag\square\triangle\ddag}$          & $\textbf{0.587}^{\star\ast\dag\square\triangle\ddag}$ & $0.478^{\star\dag\square\triangle}$          & $0.573^{\star\ast\dag\square\triangle}$          \\
                       & FT-MTL w/ PCGrad      & $0.366^{\star}$          & 0.457          & $0.468^{\star\dag\square\triangle}$          & 0.542          & $0.449^{\square\triangle}$          & 0.515          \\
                       & Dense MERIT           & $0.376^{\star}$          & $0.528^{\star\ast\ddag+}$          & $0.474^{\star\dag\square\triangle+}$          & $0.577^{\star\ast\dag\square\triangle+}$          & $0.476^{\star\dag\square\triangle+}$          & $0.570^{\star\ast\dag\square\triangle+}$          \\
                       & Sparse MERIT          & $0.379^{\star+}$          & $0.520^{\star+}$          & $\textbf{0.489}^{\star\ast\dag\square\triangle\ddag\circ+\diamond}$ & $0.586^{\star\ast\dag\square\triangle\ddag+}$          & $\textbf{0.492}^{\star\ast\dag\square\triangle\ddag\circ+\diamond}$ & $\textbf{0.579}^{\star\ast\dag\square\triangle\ddag+}$ \\ \midrule
\multirow{8}{*}{0 dB}  & Original              & 0.416          & 0.553          & 0.508          & 0.598          & 0.511          & 0.600          \\
                       & FT-M                  & $0.520^{\star}$          & $0.597^{\star}$          & $0.543^{\star}$          & 0.610          & $0.540^{\star}$          & 0.608          \\
                       & SE-P w/ BSSE-SE \cite{hung_2022_1}                  & $\textbf{0.557}^{\star\ast}$ & $0.624^{\star\ast}$ & $0.534^{\star}$          & 0.607          & $0.521^{\star}$          & 0.597          \\
                       & SE-P w/ SGMSE+ \cite{richter_2023_1}                  & $0.539^{\star\ast}$ & $0.614^{\star\ast}$ &    $0.536^{\star}$    &  $0.613^{\star}$   &   $0.521^{\star}$    &  0.599         \\
                       & SE-P w/ SEMamba \cite{chao_2024_1}                  & $0.556^{\star\ast\square}$ & $\textbf{0.640}^{\star\ast\dag\square}$ &   $0.535^{\star}$     &   $0.620^{\star\ast\dag\square}$   &   $0.522^{\star}$  &   $0.608^{\star\dag}$   \\
                       & FT-MTL \cite{tzeng_2025_1}                & $0.525^{\star\ast}$          & $0.607^{\star\ast}$          & $0.545^{\star\dag\square\triangle}$          & $0.625^{\star\ast\dag\square\triangle}$          & $0.545^{\star\ast\dag\square\triangle}$          & $0.622^{\star\ast\dag\square\triangle}$          \\
                       & FT-MTL w/ Uncertainty & $0.527^{\star\ast}$          & $0.613^{\star\ast\ddag}$          & $0.552^{\star\ast\dag\square\triangle\ddag}$          & $\textbf{0.635}^{\star\ast\dag\square\triangle\ddag}$ & $0.546^{\star\ast\dag\square\triangle}$          & $0.628^{\star\ast\dag\square\triangle\ddag}$          \\
                       & FT-MTL w/ PCGrad      & $0.504^{\star}$          & $0.569^{\star}$          & $0.538^{\star}$          & 0.599          & $0.525^{\star}$          & 0.584          \\
                       & Dense MERIT           & $0.529^{\star\ast+}$          & $0.618^{\star\ast\ddag+}$          & $0.551^{\star\ast\dag\square\triangle+}$          & $0.632^{\star\ast\dag\square\triangle\ddag+}$          & $0.545^{\star\ast\dag\square\triangle+}$          & $0.626^{\star\ast\dag\square\triangle+}$          \\
                       & Sparse MERIT          & $0.527^{\star\ast+}$          & $0.615^{\star\ast\ddag+}$          & $\textbf{0.556}^{\star\ast\dag\square\triangle\ddag+}$ & $0.634^{\star\ast\dag\square\triangle\ddag+}$          & $\textbf{0.554}^{\star\ast\dag\square\triangle\ddag\circ+\diamond}$ & $\textbf{0.631}^{\star\ast\dag\square\triangle\ddag+}$ \\ \midrule
\multirow{8}{*}{5 dB}  & Original              & 0.508          & 0.612          & 0.554          & 0.629          & 0.548          & 0.627          \\
                       & FT-M                  & $0.557^{\star}$          & $0.623^{\star}$          & $0.564^{\star}$          & 0.627          & $0.562^{\star}$          & 0.626          \\
                       & SE-P w/ BSSE-SE \cite{hung_2022_1}                  & $0.572^{\star\ast}$ & $0.636^{\star}$          & $0.564^{\star}$          & 0.629          & $0.557^{\star}$          & 0.625          \\
                       & SE-P w/ SGMSE+ \cite{richter_2023_1}                  & $0.570^{\star\ast}$ &   $0.635^{\star\ast}$      &   $0.564^{\star}$   &  0.630    &   $0.558^{\star}$    &   0.625     \\
                       & SE-P w/ SEMamba \cite{chao_2024_1}                  & $\textbf{0.579}^{\star\ast\dag\square}$ &  $\textbf{0.653}^{\star\ast\dag\square}$    &    $0.568^{\star}$    &   $0.644^{\star\ast\dag\square}$    &    $0.558^{\star}$    &    $0.637^{\star\ast\dag\square}$     \\
                       & FT-MTL \cite{tzeng_2025_1}                & $0.562^{\star\ast}$          & $0.635^{\star\ast}$          & $0.565^{\star}$          & $0.640^{\star\ast\square}$          & $0.567^{\star\ast\dag\square\triangle}$          & $0.640^{\star\ast\dag\square}$          \\
                       & FT-MTL w/ Uncertainty & $0.566^{\star\ast}$          & $0.639^{\star\ast}$          & $0.574^{\star\ast\dag\square\triangle\ddag}$          & $0.651^{\star\ast\dag\square\triangle\ddag}$          & $0.568^{\star\ast\dag\square\triangle}$          & $0.645^{\star\ast\dag\square\triangle\ddag}$          \\
                       & FT-MTL w/ PCGrad      & $0.553^{\star}$          & 0.611          & $0.564^{\star}$          & 0.622          & $0.561^{\star}$          & 0.616          \\
                       & Dense MERIT           & $0.572^{\star\ast\ddag\circ+}$ & $0.646^{\star\ast\square\ddag\circ+}$ & $0.576^{\star\ast\dag\square\triangle\ddag+}$          & $\textbf{0.652}^{\star\ast\dag\square\triangle\ddag+}$ & $0.572^{\star\ast\dag\square\triangle\ddag+}$          & $0.648^{\star\ast\dag\square\triangle\ddag+}$          \\
 & Sparse MERIT & $0.569^{\star\ast\ddag+}$                   & $0.644^{\star\ast\ddag+}$                   & $\textbf{0.578}^{\star\ast\dag\square\triangle\ddag+}$        & $\textbf{0.652}^{\star\ast\dag\square\triangle\ddag+}$       & $\textbf{0.577}^{\star\ast\dag\square\triangle\ddag\circ+}$     & $\textbf{0.651}^{\star\ast\dag\square\triangle\ddag\circ+}$    \\ \midrule
\multirow{8}{*}{10 dB} & Original              & 0.553          & 0.637          & 0.570          & 0.639          & 0.565          & 0.638          \\
                       & FT-M                  & $0.569^{\star}$          & 0.633          & 0.572          & 0.634          & 0.571          & 0.634          \\
                       & SE-P w/ BSSE-SE \cite{hung_2022_1}                  & $0.576^{\star}$          & 0.638          & 0.574          & 0.636          & 0.570          & 0.633          \\
                       & SE-P w/ SGMSE+ \cite{richter_2023_1}                  &   $0.575^{\star}$    & 0.637       &    0.573    &   0.637   &    0.570   &   0.633    \\
                       & SE-P w/ SEMamba \cite{chao_2024_1}                  &  $0.581^{\star\ast\dag}$    &   $0.654^{\star\ast\dag\square}$    &    $0.580^{\star\ast}$   &    $0.652^{\star\ast\dag\square}$   &    $0.574^{\star}$     &    $0.649^{\star\ast\dag\square}$   \\
                       & FT-MTL \cite{tzeng_2025_1}                & $0.573^{\star}$          & $0.644^{\ast}$          & 0.572          & $0.644^{\ast}$          & $0.571^{\star}$          & $0.643^{\ast\square}$          \\
                       & FT-MTL w/ Uncertainty & $0.578^{\star\ast\ddag}$          & $0.649^{\star\ast\dag\square}$          & $0.582^{\star\ast\dag\square\ddag}$          & $0.656^{\star\ast\dag\square\triangle\ddag}$          & $0.578^{\star\ast\dag\square\ddag}$          & $0.652^{\star\ast\dag\square\ddag}$          \\
                       & FT-MTL w/ PCGrad      & $0.570^{\star}$          & 0.626          & 0.574          & 0.631          & $0.574^{\star}$          & 0.629          \\
                       & Dense MERIT           & $0.582^{\star\ast\square\ddag+}$          & $0.655^{\star\ast\dag\square\ddag\circ+}$          & $0.584^{\star\ast\dag\square\ddag+}$          & $0.657^{\star\ast\dag\square\triangle\ddag+}$          & $0.582^{\star\ast\dag\square\triangle\ddag\circ+}$          & $0.654^{\star\ast\dag\square\ddag+}$          \\
 & Sparse MERIT & $\textbf{0.585}^{\star\ast\dag\ddag+}$          & $\textbf{0.658}^{\star\ast\dag\square\ddag\circ+}$          & $\textbf{0.586}^{\star\ast\dag\square\triangle\ddag+}$        & $\textbf{0.660}^{\star\ast\dag\square\triangle\ddag+}$       & $\textbf{0.583}^{\star\ast\dag\square\triangle\ddag}$     & $\textbf{0.658}^{\star\ast\dag\square\triangle\ddag\circ+}$    \\ \bottomrule
\end{tabular}
}
\end{center}
\label{table:ser_performance}
\end{table*}

We evaluate SER performance using both F1-macro and F1-micro scores across four SNR levels, under both seen and unseen noise conditions. Each method is trained using four different random seeds, and the test set is divided into five non-overlapping subsets per condition, yielding 20 evaluation scores per method (4 runs $\times$ 5 test sets). These scores are used to compute average performance and conduct statistical comparisons. We apply one-tailed Welch’s t-tests to compare each method against all other baselines. Statistical significance is determined at a threshold of $p \leq 0.05$. Significance markers in Table \ref{table:ser_performance} indicate whether a method outperforms a given baseline, with symbol definitions provided in the table caption.

Our proposed Sparse MERIT framework achieves statistically significant improvements over baselines under low-SNR and unseen noisy conditions, demonstrating strong robustness and generalization across diverse acoustic scenarios. At -5 dB, the most challenging condition, Sparse MERIT yields an F1-macro improvement of 3.6\% over fine-tuning directly on noisy speech (FT-M), 12.4\% over the SE pre-processing baseline using BSSE-SE (SE-P w/ BSSE-SE), 6.5\% over the SE-P w/ SGMSE+, 10.9\% over the SE-P w/ SEMamba, and 3.8\% over the naive MTL setup (FT-MTL) on the Freesound-contaminated test set. Similar trends are observed on the DNS-contaminated test set, where Sparse MERIT outperforms FT-M, SE-P w/ BSSE-SE, SE-P w/ SGMSE+, SE-P w/ SEMamba, and FT-MTL by 3.8\%, 11.6\%, 11.3\%, 13.6\%, and 2.9\%, respectively. While SE-P performs relatively well under seen noise conditions, its performance degrades notably under unseen noise, highlighting its limited generalization. In contrast, Sparse MERIT maintains robust performance across both unseen noisy conditions. Furthermore, under high-SNR conditions, our method performs comparably or better than SE-P even on the seen CRSS test set, suggesting that it avoids the artifacts and loss of emotional nuance introduced by front-end enhancement methods applied to minimally corrupted signals, as reported in prior work \cite{chen_2023_1}.

Beyond baseline comparisons, we also evaluate Sparse MERIT against other MTL strategies designed to mitigate task conflicts (FT-MTL w/ Uncertainty and FT-MTL w/ PCGrad). Uncertainty-based loss weighting yields promising results and generally outperforms standard FT-MTL. However, PCGrad fails to show consistent benefits in our setting and underperforms naive MTL, indicating its limited utility in this task combination. When we compare architectural variants of our approach, Sparse MERIT outperforms Dense MERIT, achieving F1-macro score gains of 1.5\% using Freesound noises and 1.6\% using DNS noises. These results suggest that Top-1 expert routing, by assigning each frame to a single expert, encourages more focused and stable specialization, leading to better generalization and efficiency under noisy conditions.

\subsection{Speech Enhancement}

We evaluate the SE performance of each method across three noisy conditions and four SNR levels (–5 dB, 0 dB, 5 dB, and 10 dB). Each model is trained using a fixed random seed to ensure consistency. We report six widely used objective metrics to assess SE quality: PESQ (Perceptual Evaluation of Speech Quality), CSIG (Mean Opinion Score of signal distortion), CBAK (Mean Opinion Score of background noise intrusiveness), COVL (Mean Opinion Score of overall quality), SSNR (Segmental SNR), and STOI (Short-Time Objective Intelligibility). These metrics offer a comprehensive assessment of both the perceptual quality and intelligibility of the enhanced speech across varying noise levels and conditions.

As shown in Table \ref{table:se_performance}, the fine-tuned BSSE-SE model achieves the best performance across all four SNR levels under the seen CRSS noise condition. However, when tested on unseen noise conditions such as the Freesound-contaminated test set, Sparse MERIT consistently outperforms all baselines across SNR levels on most objective metrics, with the only exception that its PESQ is slightly lower than SEMamba. This is expected because SEMamba incorporates a PESQ-oriented discriminator, which explicitly optimizes for PESQ. Focusing on the most challenging -5 dB condition, Sparse MERIT shows a 1.8\% drop in PESQ, but 14.1\% improvement in CSIG, 5.3\% in CBAK, 9.3\% in COVL, 12.1\% in SSNR, and a 20.0\% improvement in STOI compared to the BSSE-SE model. Against the naive multi-task learning baseline (FT-MTL), Sparse MERIT shows 0.9\% higher PESQ, 6.1\% higher CSIG, 2.6\% higher CBAK, 4.5\% higher COVL, 8.6\% higher SSNR, and 3.1\% higher STOI.

On the DNS-contaminated test set, Sparse MERIT again demonstrates superior performance at –5, 0, and 5 dB, and performs comparably to other methods at 10 dB. At –5 dB, compared to the BSSE-SE SE-only model, it yields equal PESQ, but achieves 15.1\% higher CSIG, 7.7\% higher CBAK, 10.3\% higher COVL, 44.4\% higher SSNR, and 19.6\% higher STOI. Relative to FT-MTL, it improves 1.8\% on PESQ, 7.6\% on CSIG, 4.0\% on CBAK, 6.0\% on COVL, 31.5\% on SSNR, and 4.7\% on STOI. These results demonstrate that Sparse MERIT not only generalizes better across unseen noise but also enhances intelligibility and perceptual quality under extremely low-SNR conditions.

While uncertainty loss weighting improves SER performance over the FT-MTL baseline, it does not yield noticeable gains for SE. In contrast, PCGrad does not improve performance on either task, yielding results that are comparable to or worse than those of naive multi-task learning. Sparse MERIT, on the other hand, demonstrates consistent benefits across both tasks. For SER, Sparse MERIT achieves superior performance, likely due to its more focused expert routing. For SE, both the dense and sparse MERIT variants deliver strong and comparable results, indicating that the expert-based integration mechanism supports robust enhancement regardless of the routing strategy.

\begin{table*}[]
\caption{SE performance of the proposed method (Sparse MERIT), all baseline models, and the unprocessed noisy recordings.}
\centering
\resizebox{\linewidth}{!}{
\begin{tabular}{cccccccc|cccccc|cccccc}
\hline
\multicolumn{8}{c|}{CRSS-4ENGLISH-14 (Seen)} &
  \multicolumn{6}{c|}{Freesound (Unseen)} &
  \multicolumn{6}{c}{DNS (Unseen)} \\ \hline
SNR &
  Model &
  PESQ &
  CSIG &
  CBAK &
  COVL &
  SSNR &
  STOI &
  PESQ &
  CSIG &
  CBAK &
  COVL &
  SSNR &
  STOI &
  PESQ &
  CSIG &
  CBAK &
  COVL &
  SSNR &
  STOI \\ \hline
\multirow{9}{*}{-5 dB} &
  Noisy &
  1.08 &
  1.77 &
  1.37 &
  1.34 &
  -5.38 &
  0.49 &
  1.09 &
  2.14 &
  1.48 &
  1.55 &
  -5.32 &
  0.63 &
  1.11 &
  1.99 &
  1.68 &
  1.50 &
  -2.58 &
  0.65 \\
 &
 BSSE-SE \cite{hung_2022_1}&
  \textbf{1.29} &
  \textbf{2.77} &
  \textbf{2.19} &
  \textbf{2.02} &
  \textbf{2.06} &
  \textbf{0.68} &
  \textbf{1.14} &
  1.99 &
  1.52 &
  1.50 &
  -4.72 &
  0.55 &
  \textbf{1.15} &
  1.85 &
  1.69 &
  1.45 &
  -2.23 &
  0.56 \\
 &
  SGMSE+ \cite{richter_2023_1}&
  1.08 &
  1.76 &
  1.42 &
  1.32 &
  -4.01 &
  0.47 &
  1.09 &
  1.98 &
  1.46 &
  1.47 &
  -5.11 &
  0.59 &
  1.11 &
  1.79 &
  1.62 &
  1.40 &
  -2.81 &
  0.59 \\
 &
  SEMamba \cite{chao_2024_1}&
  1.17 &
  2.18 &
  1.71 &
  1.60 &
  -2.02 &
  0.57 &
  1.13 &
  1.99 &
  1.29 &
  1.49 &
  -8.16 &
  0.59 &
  1.13 &
  1.75 &
  1.47 &
  1.39 &
  -5.13 &
  0.57 \\
 &
  FT-MTL \cite{tzeng_2025_1}&
  1.16 &
  2.48 &
  1.91 &
  1.78 &
  -0.01 &
  0.60 &
  1.11 &
  2.14 &
  1.56 &
  1.57 &
  -4.54 &
  0.64 &
  1.13 &
  1.98 &
  1.75 &
  1.51 &
  -1.81 &
  0.64 \\
 &
  FT-MTL w/ Uncertainty &
  1.17 &
  2.52 &
  1.94 &
  1.81 &
  0.20 &
  0.62 &
  1.11 &
  2.17 &
  1.55 &
  1.58 &
  -4.59 &
  0.64 &
  1.13 &
  2.00 &
  1.75 &
  1.52 &
  -1.84 &
  0.64 \\
 &
  FT-MTL w/ PCGrad &
  1.16 &
  2.46 &
  1.94 &
  1.77 &
  0.25 &
  0.60 &
  1.11 &
  2.18 &
  1.55 &
  1.59 &
  -4.63 &
  0.64 &
  1.13 &
  2.00 &
  1.74 &
  1.52 &
  -1.90 &
  0.64 \\
 &
  Dense MERIT &
  1.14 &
  2.38 &
  1.83 &
  1.71 &
  -0.69 &
  0.58 &
  1.12 &
  2.23 &
  1.59 &
  1.62 &
  -4.22 &
  0.65 &
  1.14 &
  2.06 &
  1.78 &
  1.56 &
  -1.51 &
  0.65 \\
 &
  Sparse MERIT &
  1.13 &
  2.32 &
  1.81 &
  1.67 &
  -0.85 &
  0.57 &
  1.12 &
  \textbf{2.27} &
  \textbf{1.60} &
  \textbf{1.64} &
  \textbf{-4.15} &
  \textbf{0.66} &
  \textbf{1.15} &
  \textbf{2.13} &
  \textbf{1.82} &
  \textbf{1.60} &
  \textbf{-1.24} &
  \textbf{0.67} \\ \hline
\multirow{9}{*}{0dB} &
  Noisy &
  1.10 &
  2.16 &
  1.68 &
  1.57 &
  -2.35 &
  0.62 &
  1.15 &
  2.50 &
  1.79 &
  1.79 &
  -2.28 &
  0.73 &
  1.16 &
  2.34 &
  1.98 &
  1.72 &
  0.52 &
  0.74 \\
 &
 BSSE-SE \cite{hung_2022_1}&
  \textbf{1.76} &
  \textbf{3.51} &
  \textbf{2.72} &
  \textbf{2.66} &
  \textbf{5.24} &
  \textbf{0.83} &
  1.26 &
  2.69 &
  1.99 &
  1.95 &
  -0.51 &
  0.75 &
  1.27 &
  2.48 &
  2.15 &
  1.85 &
  1.99 &
  0.72 \\
 &
  SGMSE+ \cite{richter_2023_1}&
  1.21 &
  2.57 &
  2.03 &
  1.85 &
  1.15 &
  0.68 &
  1.16 &
  2.45 &
  1.80 &
  1.77 &
  -1.93 &
  0.71 &
  1.16 &
  2.22 &
  1.95 &
  1.65 &
  0.37 &
  0.70 \\
 &
  SEMamba \cite{chao_2024_1}&
  1.70 &
  3.24 &
  2.45 &
  2.48 &
  2.08 &
  0.79 &
  \textbf{1.32} &
  2.59 &
  1.64 &
  1.92 &
  -6.03 &
  0.74 &
  1.26 &
  2.28 &
  1.77 &
  1.74 &
  -3.35 &
  0.72 \\
 &
  FT-MTL \cite{tzeng_2025_1}&
  1.59 &
  3.34 &
  2.55 &
  2.48 &
  4.35 &
  0.81 &
  1.23 &
  2.68 &
  1.96 &
  1.94 &
  -0.80 &
  0.76 &
  1.24 &
  2.51 &
  2.16 &
  1.86 &
  2.02 &
  0.76 \\
 &
  FT-MTL w/ Uncertainty &
  1.59 &
  3.34 &
  2.56 &
  2.48 &
  4.38 &
  0.81 &
  1.23 &
  2.70 &
  1.95 &
  1.94 &
  -0.87 &
  0.77 &
  1.24 &
  2.51 &
  2.15 &
  1.86 &
  1.95 &
  0.76 \\
 &
  FT-MTL w/ PCGrad &
  1.55 &
  3.28 &
  2.52 &
  2.43 &
  4.22 &
  0.80 &
  1.23 &
  2.71 &
  1.94 &
  1.94 &
  -0.95 &
  0.76 &
  1.23 &
  2.52 &
  2.14 &
  1.86 &
  1.88 &
  0.76 \\
 &
  Dense MERIT &
  1.57 &
  3.31 &
  2.52 &
  2.45 &
  4.13 &
  0.80 &
  1.28 &
  2.84 &
  2.04 &
  2.04 &
  -0.09 &
  \textbf{0.78} &
  1.29 &
  2.68 &
  2.24 &
  1.97 &
  2.75 &
  \textbf{0.78} \\
 &
  Sparse MERIT &
  1.48 &
  3.18 &
  2.44 &
  2.33 &
  3.70 &
  0.78 &
  1.29 &
  \textbf{2.86} &
  \textbf{2.06} &
  \textbf{2.06} &
  \textbf{0.12} &
  \textbf{0.78} &
  \textbf{1.31} &
  \textbf{2.73} &
  \textbf{2.27} &
  \textbf{2.01} &
  \textbf{2.98} &
  \textbf{0.78} \\ \hline
\multirow{9}{*}{5dB} &
  Noisy &
  1.20 &
  2.61 &
  2.06 &
  1.88 &
  1.20 &
  0.74 &
  1.29 &
  2.91 &
  2.16 &
  2.09 &
  1.25 &
  0.82 &
  1.29 &
  2.75 &
  2.35 &
  2.02 &
  4.06 &
  0.82 \\
 &
 BSSE-SE \cite{hung_2022_1}&
  2.31 &
  \textbf{4.04} &
  \textbf{3.19} &
  \textbf{3.21} &
  \textbf{8.03} &
  \textbf{0.89} &
  1.66 &
  3.38 &
  2.59 &
  2.53 &
  4.39 &
  0.86 &
  1.63 &
  3.23 &
  2.73 &
  2.44 &
  6.74 &
  0.84 \\
 &
  SGMSE+ \cite{richter_2023_1}&
  1.72 &
  3.45 &
  2.76 &
  2.60 &
  6.18 &
  0.84 &
  1.36 &
  2.97 &
  2.24 &
  2.16 &
  1.95 &
  0.81 &
  1.32 &
  2.73 &
  2.36 &
  2.02 &
  4.11 &
  0.80 \\
 &
  SEMamba \cite{chao_2024_1}&
  \textbf{2.38} &
  3.93 &
  2.95 &
  3.19 &
  3.78 &
  0.87 &
  \textbf{1.83} &
  3.36 &
  2.22 &
  2.60 &
  -2.43 &
  0.85 &
  1.63 &
  3.00 &
  2.23 &
  2.32 &
  -0.65 &
  0.83 \\
 &
  FT-MTL \cite{tzeng_2025_1}&
  2.16 &
  3.92 &
  3.10 &
  3.07 &
  7.74 &
  0.88 &
  1.59 &
  3.32 &
  2.51 &
  2.46 &
  3.80 &
  0.86 &
  1.55 &
  3.16 &
  2.68 &
  2.37 &
  6.44 &
  0.85 \\
 &
  FT-MTL w/ Uncertainty &
  2.16 &
  3.92 &
  3.10 &
  3.07 &
  7.76 &
  \textbf{0.89} &
  1.59 &
  3.33 &
  2.52 &
  2.47 &
  3.84 &
  0.86 &
  1.55 &
  3.15 &
  2.67 &
  2.36 &
  6.40 &
  0.85 \\
 &
  FT-MTL w/ PCGrad &
  2.12 &
  3.88 &
  3.07 &
  3.03 &
  7.63 &
  0.88 &
  1.57 &
  3.33 &
  2.49 &
  2.46 &
  3.61 &
  0.86 &
  1.54 &
  3.16 &
  2.67 &
  2.37 &
  6.37 &
  0.85 \\
 &
  Dense MERIT &
  2.14 &
  3.91 &
  3.08 &
  3.06 &
  7.66 &
  0.88 &
  1.71 &
  3.50 &
  2.64 &
  2.63 &
  4.67 &
  \textbf{0.87} &
  1.70 &
  3.40 &
  2.83 &
  2.57 &
  7.46 &
  \textbf{0.87} \\
 &
  Sparse MERIT &
  1.97 &
  3.74 &
  2.96 &
  2.89 &
  7.19 &
  0.87 &
  1.74 &
  \textbf{3.53} &
  \textbf{2.68} &
  \textbf{2.65} &
  \textbf{5.07} &
  \textbf{0.87} &
  \textbf{1.72} &
  \textbf{3.41} &
  \textbf{2.84} &
  \textbf{2.59} &
  \textbf{7.49} &
  \textbf{0.87} \\ \hline
\multirow{9}{*}{10 dB} &
  Noisy &
  1.42 &
  3.09 &
  2.50 &
  2.26 &
  5.15 &
  0.83 &
  1.57 &
  3.36 &
  2.60 &
  2.48 &
  5.17 &
  0.88 &
  1.53 &
  3.20 &
  2.77 &
  2.38 &
  7.90 &
  0.88 \\
 &
  BSSE-SE \cite{hung_2022_1}&
  \textbf{2.81} &
  \textbf{4.46} &
  \textbf{3.63} &
  \textbf{3.68} &
  \textbf{10.82} &
  \textbf{0.93} &
  2.28 &
  4.03 &
  3.23 &
  3.19 &
  8.99 &
  \textbf{0.92} &
  2.18 &
  3.88 &
  3.32 &
  3.07 &
  10.98 &
  0.91 \\
 &
  SGMSE+ \cite{richter_2023_1}&
  2.30 &
  4.04 &
  3.29 &
  3.21 &
  9.44 &
  0.91 &
  1.79 &
  3.57 &
  2.80 &
  2.70 &
  6.41 &
  0.89 &
  1.66 &
  3.31 &
  2.86 &
  2.51 &
  8.13 &
  0.88 \\
 &
  SEMamba \cite{chao_2024_1}&
  2.75 &
  4.25 &
  3.16 &
  3.54 &
  4.15 &
  0.91 &
  \textbf{2.51} &
  4.04 &
  2.82 &
  \textbf{3.31} &
  0.91 &
  0.90 &
  2.23 &
  3.72 &
  2.74 &
  3.01 &
  1.79 &
  0.89 \\
 &
  FT-MTL \cite{tzeng_2025_1}&
  2.68 &
  4.36 &
  3.55 &
  3.56 &
  10.67 &
  0.92 &
  2.20 &
  3.97 &
  3.16 &
  3.12 &
  8.48 &
  \textbf{0.92} &
  2.10 &
  3.81 &
  3.26 &
  2.99 &
  10.67 &
  0.91 \\
 &
  FT-MTL w/ Uncertainty &
  2.68 &
  4.35 &
  3.55 &
  3.56 &
  10.68 &
  0.92 &
  2.22 &
  3.99 &
  3.19 &
  3.14 &
  8.68 &
  \textbf{0.92} &
  2.10 &
  3.80 &
  3.26 &
  2.99 &
  10.67 &
  0.91 \\
 &
  FT-MTL w/ PCGrad &
  2.64 &
  4.33 &
  3.53 &
  3.53 &
  10.58 &
  0.92 &
  2.18 &
  3.97 &
  3.14 &
  3.11 &
  8.29 &
  \textbf{0.92} &
  2.09 &
  3.81 &
  3.25 &
  2.99 &
  10.64 &
  0.91 \\
 &
  Dense MERIT &
  2.65 &
  4.34 &
  3.54 &
  3.54 &
  10.61 &
  0.92 &
  2.32 &
  4.10 &
  3.26 &
  3.24 &
  9.04 &
  \textbf{0.92} &
  \textbf{2.29} &
  \textbf{4.01} &
  \textbf{3.41} &
  \textbf{3.19} &
  \textbf{11.46} &
  \textbf{0.92} \\
 &
  Sparse MERIT &
  2.47 &
  4.18 &
  3.41 &
  3.36 &
  10.13 &
  0.91 &
  2.33 &
  \textbf{4.11} &
  \textbf{3.29} &
  3.26 &
  \textbf{9.34} &
  \textbf{0.92} &
  2.28 &
  3.99 &
  3.39 &
  3.18 &
  11.25 &
  \textbf{0.92} \\ \hline
\end{tabular}
}
\label{table:se_performance}
\end{table*}

\subsection{Impact of Expert Network Size}
In this section, we investigate how varying the number of experts affects both SE and SER performance. We evaluate models with 1, 3, 5, 7, and 9 experts using a fixed random seed to ensure a fair comparison. For the SE evaluation, we use SSNR, as it provides a more neutral assessment of enhancement quality. In contrast to perceptual metrics such as PESQ, STOI, CSIG, COVL, and CBAK, which emphasize intelligibility or human-perceived quality, SSNR does not inherently favor clearer or more intelligible speech. This makes it more suitable in our case, where we aim to evaluate enhancement quality without biasing toward intelligibility or emotional nuance.

As shown in Table \ref{table:experts_size}, performance trends differ between the two tasks. SER performance peaks when using 3 experts, whereas SE performance generally improves with more experts, reaching its highest SSNR at 5.

These findings suggest that, unlike in large language models where increasing expert count often improves performance \cite{fedus_2022_1}, multi-task SE and SER learning does not exhibit this pattern. Given that our primary goal is to enhance SER robustness, we adopt three experts, as they consistently deliver the best SER results across unseen noise conditions.

% \begin{figure}[!t]
% \includegraphics[width=0.45\textwidth]{Figures/experts_size_dev.png}
% \centering
% \caption{SER and SE performance of models with varying numbers of experts, evaluated on the development set. Each point represents a single model configuration, with the number of experts indicated next to the corresponding marker.}
% \label{fig:experts_size}
% \end{figure}

\begin{table}[]
\centering
\caption{SER and SE performance of models with varying numbers of experts, evaluated on the development set.}
\begin{tabular}{cccccc}
\hline
Number of experts \# & 1     & 3              & 5             & 7     & 9     \\ \hline
F1-macro             & 0.564 & \textbf{0.583} & 0.577         & 0.573 & 0.568 \\
SSNR (dB)            & 7.14  & 7.35           & \textbf{7.62} & 7.58  & 7.57  \\ \hline
\end{tabular}
\label{table:experts_size}
\end{table}

\subsection{Analysis of Gating Behavior}
To better understand how the MoE mechanism operates under different acoustic and emotional conditions, we analyzed three aspects of gating behavior: switching dynamics, agreement between SE and SER gates, and expert usage distributions across SNR levels and emotion classes. The reported values are aggregated from all three testing sets.

\subsubsection{Switching dynamics}
The switching rate quantifies how frequently the gating function changes its expert selection across consecutive frames. A higher switching rate indicates less temporal stability and greater responsiveness to acoustic variations. As shown in Table \ref{table:gate_behavior_snr}, switching rates slightly decrease as SNR increases, meaning the gate becomes more stable when the input is less noisy. Conversely, at low SNR (–5 dB), both SE and SER exhibited more frequent switching, consistent with the need to adapt to challenging acoustic conditions. Examining emotion classes in Table \ref{table:gate_behavior_emotion}, Sad and Neutral utterances required more switching than Angry or Happy, suggesting that their acoustic profiles prompted more dynamic expert selection.

\subsubsection{Agreement between SE and SER gates}
Agreement measures the proportion of frames where the SE and SER tasks select the same expert. Across SNR levels, agreement values were remarkably stable, indicating that noise conditions did not strongly affect the extent of shared expert usage as shown in Table \ref{table:gate_behavior_snr}. Clear differences emerged across emotions in Table \ref{table:gate_behavior_emotion}, where Angry utterances showed the lowest agreement and Happy utterances achieved the highest agreement. This suggests that emotional content, rather than noise, primarily drives divergence or convergence between SE and SER gating.

\begin{table}[]
\centering
\caption{Switch rate and SE–SER agreement across SNR levels.}
\begin{tabular}{cllll}
\hline
SNR & \multicolumn{1}{c}{-5 dB} & \multicolumn{1}{c}{0 dB} & \multicolumn{1}{c}{5 dB} & \multicolumn{1}{c}{10 dB} \\ \hline
SE switch  & \textbf{0.316} & 0.305 & 0.294 & 0.290 \\
SER switch & \textbf{0.314} & 0.303 & 0.297 & 0.298 \\
Agreement  & \textbf{0.406} & 0.396 & 0.397 & 0.401 \\ \hline
\end{tabular}
\label{table:gate_behavior_snr}
\end{table}

\begin{table}[]
\centering
\caption{Switch rate and SE–SER agreement across emotion labels.}
\begin{tabular}{ccccc}
\hline
Label    & Angry & Sad   & Happy & Neutral \\ \hline
SE switch  & 0.297 & 0.299 & 0.288 & \textbf{0.311}   \\
SER switch & 0.292 & \textbf{0.317} & 0.300 & 0.306   \\
Agreement  & 0.314 & 0.338 & \textbf{0.448} & 0.401   \\ \hline
\end{tabular}
\label{table:gate_behavior_emotion}
\end{table}

\subsubsection{Expert usage distributions}
\begin{figure}[!t]
\includegraphics[width=0.5\textwidth]{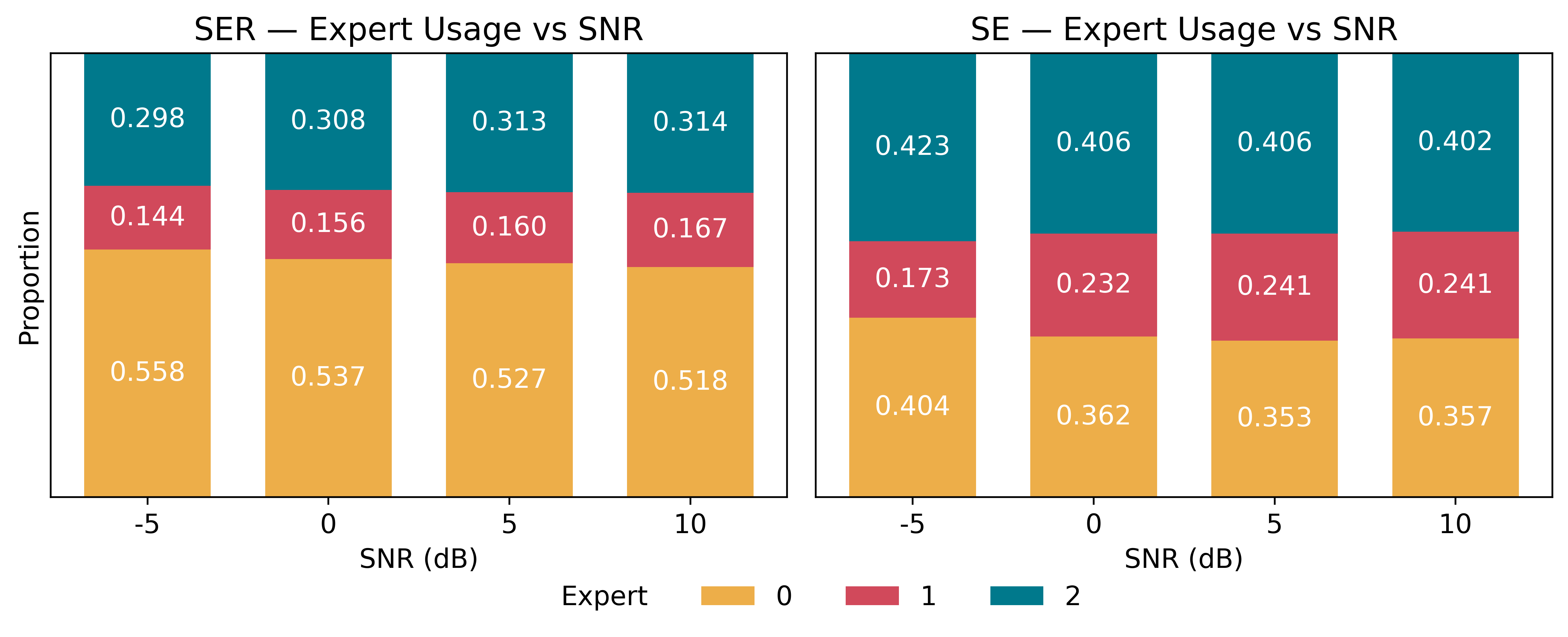}
\centering
\caption{Expert usage distributions across SNR conditions.}
\label{fig:emotion_expert_usage_snr}
\end{figure}

\begin{figure}[!t]
\includegraphics[width=0.5\textwidth]{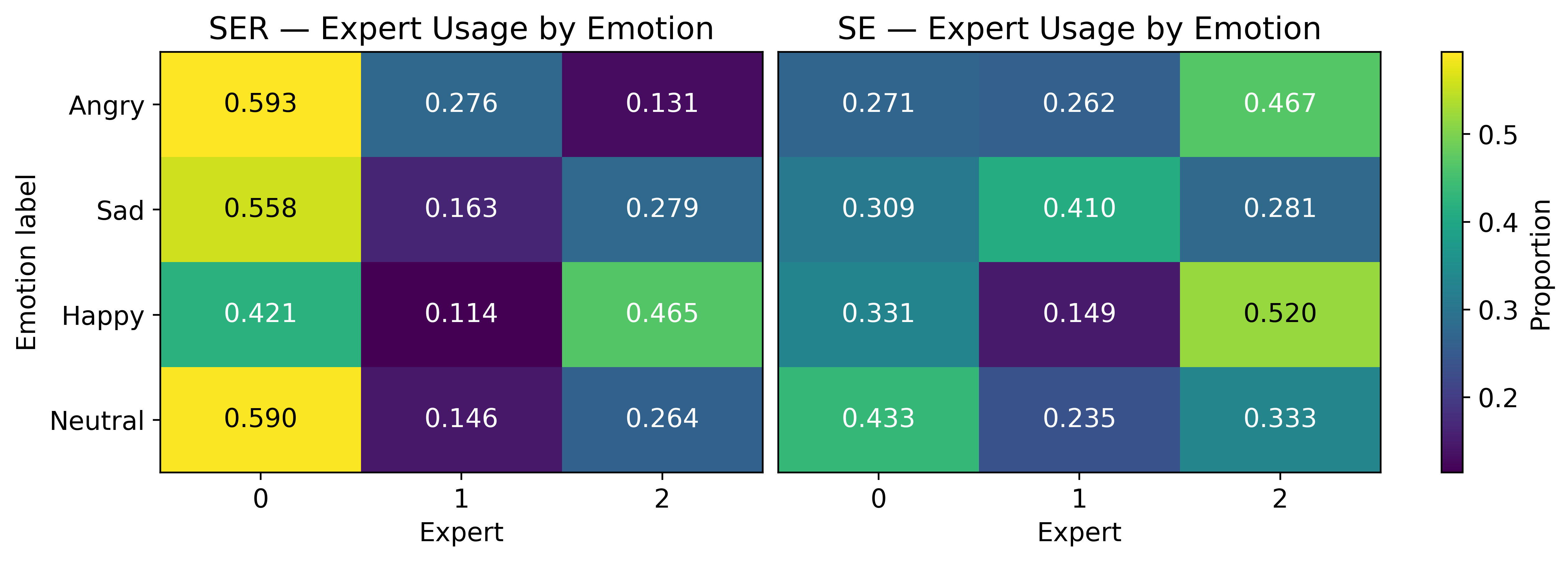}
\centering
\caption{Frame-level expert usage distributions across emotion classes.}
\label{fig:emotion_expert_usage_emotion}
\end{figure}

Expert usage reflects the long-term allocation of frames to each expert. For SER, Expert 0 dominated across all conditions, but its contribution decreased with increasing SNR, while Experts 1 and 2 became more utilized as illustrated in Fig. \ref{fig:emotion_expert_usage_snr}. This indicates that under cleaner conditions, the gate distributes its reliance more evenly. For SE, the trend was less pronounced: Expert 0 and 2 usage decreased with SNR, but usage across Experts 0, 1, and 2 remained relatively steady at 0, 5, and 10 dB as shown in Fig. \ref{fig:emotion_expert_usage_snr}. Considering emotion classes in Fig. \ref{fig:emotion_expert_usage_emotion}, SER remained heavily reliant on Expert 0 overall, with Happy utterances also showing strong reliance on Expert 2. For SE, specialization was clearer: Neutral relied more on Expert 0, Sad on Expert 1, and Angry/Happy on Expert 2. These patterns suggest that while SER favors a dominant expert with some emotion-dependent variation, SE distributes responsibilities more evenly and shows stronger emotion-dependent specialization.

\subsection{Ablation Study}
\subsubsection{Effect of Expert Balancing Loss}

\begin{table*}[]
\caption{SER performance comparison of the Sparse MERIT model trained with and without the expert balancing loss.}
\begin{center}
\resizebox{0.75\linewidth}{!}{
\begin{tabular}{@{}ccccccccc@{}}
\toprule
 &
   &
   &
  \multicolumn{2}{c}{CRSS-4ENGLISH-14 (Seen)} &
  \multicolumn{2}{c}{Freesound (Unseen)} &
  \multicolumn{2}{c}{DNS (Unseen)} \\ \midrule
SNR &
  Model Architecture &
  Expert Balancing Loss &
  F1-Macro &
  F1-Micro &
  F1-Macro &
  F1-Micro &
  F1-Macro &
  F1-Micro \\ \midrule
\multirow{2}{*}{-5 dB} &
  Sparse MERIT &
  \xmark &
  0.393 &
  0.536 &
  \textbf{0.512} &
  \textbf{0.605} &
  \textbf{0.520} &
  \textbf{0.600} \\
 & Sparse MERIT & \cmark & \textbf{0.421} & \textbf{0.542} & 0.477 & 0.574 & 0.454 & 0.547 \\ \midrule
\multirow{2}{*}{0 dB} &
  Sparse MERIT &
  \xmark &
  0.549 &
  \textbf{0.634} &
  \textbf{0.574} &
  \textbf{0.650} &
  \textbf{0.575} &
  \textbf{0.649} \\
 & Sparse MERIT & \cmark & \textbf{0.552} & 0.630          & 0.559 & 0.634 & 0.552 & 0.630 \\ \midrule
\multirow{2}{*}{5 dB} &
  Sparse MERIT &
  \xmark &
  \textbf{0.588} &
  \textbf{0.660} &
  \textbf{0.596} &
  \textbf{0.668} &
  \textbf{0.596} &
  \textbf{0.667} \\
 & Sparse MERIT & \cmark & 0.581          & 0.647          & 0.589 & 0.653 & 0.579 & 0.646 \\ \midrule
\multirow{2}{*}{10 dB} &
  Sparse MERIT &
  \xmark &
  \textbf{0.603} &
  \textbf{0.673} &
  \textbf{0.607} &
  \textbf{0.676} &
  \textbf{0.603} &
  \textbf{0.673} \\
 & Sparse MERIT & \cmark & 0.592          & 0.655          & 0.598 & 0.659 & 0.591 & 0.654 \\ \bottomrule
\end{tabular}
}
\end{center}
\label{table:ser_load_balance}
\end{table*}

\begin{table}[]
\caption{SE performance comparison of the Sparse MERIT model trained with and without the expert balancing loss.}
\centering
\resizebox{0.9\linewidth}{!}{
\begin{tabular}{@{}cccccccc@{}}
\toprule
SNR &
  Method &
  PESQ &
  CSIG &
  CBAK &
  COVL &
  SSNR &
  STOI \\ \midrule
\multicolumn{8}{c}{CRSS-4ENGLISH-14 (Seen)} \\ \midrule
\multirow{2}{*}{-5 dB} &
  Sparse MERIT &
  1.13 &
  2.32 &
  1.81 &
  1.67 &
  -0.85 &
  0.57 \\
 &
  \begin{tabular}[c]{@{}c@{}}Sparse MERIT\\ w/ Expert Balancing Loss\end{tabular} &
  \textbf{1.17} &
  \textbf{2.50} &
  \textbf{1.92} &
  \textbf{1.79} &
  \textbf{0.06} &
  \textbf{0.61} \\ \midrule
\multirow{2}{*}{0 dB} &
  Sparse MERIT &
  1.48 &
  3.18 &
  2.44 &
  2.33 &
  3.70 &
  0.78 \\
 &
  \begin{tabular}[c]{@{}c@{}}Sparse MERIT\\ w/ Expert Balancing Loss\end{tabular} &
  \textbf{1.60} &
  \textbf{3.36} &
  \textbf{2.56} &
  \textbf{2.49} &
  \textbf{4.41} &
  \textbf{0.81} \\ \midrule
\multirow{2}{*}{5 dB} &
  Sparse MERIT &
  1.97 &
  3.74 &
  2.96 &
  2.89 &
  7.19 &
  0.87 \\
 &
  \begin{tabular}[c]{@{}c@{}}Sparse MERIT\\ w/ Expert Balancing Loss\end{tabular} &
  \textbf{2.16} &
  \textbf{3.93} &
  \textbf{3.10} &
  \textbf{3.08} &
  \textbf{7.75} &
  \textbf{0.89} \\ \midrule
\multirow{2}{*}{10 dB} &
  Sparse MERIT &
  2.47 &
  4.18 &
  3.41 &
  3.36 &
  10.13 &
  0.91 \\
 &
  \begin{tabular}[c]{@{}c@{}}Sparse MERIT\\ w/ Expert Balancing Loss\end{tabular} &
  \textbf{2.67} &
  \textbf{4.35} &
  \textbf{3.55} &
  \textbf{3.55} &
  \textbf{10.63} &
  \textbf{0.92} \\ \midrule
\multicolumn{8}{c}{Freesound (Unseen)} \\ \midrule
\multirow{2}{*}{-5 dB} &
  Sparse MERIT &
  \textbf{1.12} &
  \textbf{2.27} &
  \textbf{1.60} &
  \textbf{1.64} &
  \textbf{-4.15} &
  \textbf{0.66} \\
 &
  \begin{tabular}[c]{@{}c@{}}Sparse MERIT\\ w/ Expert Balancing Loss\end{tabular} &
  \textbf{1.12} &
  2.15 &
  1.56 &
  1.58 &
  -4.46 &
  0.64 \\ \midrule
\multirow{2}{*}{0 dB} &
  Sparse MERIT &
  \textbf{1.29} &
  \textbf{2.86} &
  \textbf{2.06} &
  \textbf{2.06} &
  \textbf{0.12} &
  \textbf{0.78} \\
 &
  \begin{tabular}[c]{@{}c@{}}Sparse MERIT\\ w/ Expert Balancing Loss\end{tabular} &
  1.26 &
  2.74 &
  2.00 &
  1.98 &
  -0.44 &
  0.77 \\ \midrule
\multirow{2}{*}{5 dB} &
  Sparse MERIT &
  \textbf{1.74} &
  \textbf{3.53} &
  \textbf{2.68} &
  \textbf{2.65} &
  \textbf{5.07} &
  \textbf{0.87} \\
 &
  \begin{tabular}[c]{@{}c@{}}Sparse MERIT\\ w/ Expert Balancing Loss\end{tabular} &
  1.67 &
  3.42 &
  2.60 &
  2.56 &
  4.44 &
  \textbf{0.87} \\ \midrule
\multirow{2}{*}{10 dB} &
  Sparse MERIT &
  \textbf{2.33} &
  \textbf{4.11} &
  \textbf{3.29} &
  \textbf{3.26} &
  \textbf{9.34} &
  \textbf{0.92} \\
 &
  \begin{tabular}[c]{@{}c@{}}Sparse MERIT\\ w/ Expert Balancing Loss\end{tabular} &
  2.31 &
  4.07 &
  3.25 &
  3.23 &
  9.04 &
  \textbf{0.92} \\ \midrule
\multicolumn{8}{c}{DNS (Unseen)} \\ \midrule
\multirow{2}{*}{-5 dB} &
  Sparse MERIT &
  \textbf{1.15} &
  \textbf{2.13} &
  \textbf{1.82} &
  \textbf{1.60} &
  \textbf{-1.24} &
  \textbf{0.67} \\
 &
  \begin{tabular}[c]{@{}c@{}}Sparse MERIT\\ w/ Expert Balancing Loss\end{tabular} &
  1.14 &
  2.01 &
  1.76 &
  1.53 &
  -1.78 &
  0.64 \\ \midrule
\multirow{2}{*}{0 dB} &
  Sparse MERIT &
  \textbf{1.31} &
  \textbf{2.73} &
  \textbf{2.27} &
  \textbf{2.01} &
  \textbf{2.98} &
  \textbf{0.78} \\
 &
  \begin{tabular}[c]{@{}c@{}}Sparse MERIT\\ w/ Expert Balancing Loss\end{tabular} &
  1.26 &
  2.58 &
  2.19 &
  1.91 &
  2.29 &
  0.77 \\ \midrule
\multirow{2}{*}{5 dB} &
  Sparse MERIT &
  \textbf{1.72} &
  \textbf{3.41} &
  \textbf{2.84} &
  \textbf{2.59} &
  \textbf{7.49} &
  \textbf{0.87} \\
 &
  \begin{tabular}[c]{@{}c@{}}Sparse MERIT\\ w/ Expert Balancing Loss\end{tabular} &
  1.64 &
  3.27 &
  2.76 &
  2.48 &
  6.98 &
  0.86 \\ \midrule
\multirow{2}{*}{10 dB} &
  Sparse MERIT &
  \textbf{2.28} &
  \textbf{3.99} &
  \textbf{3.39} &
  \textbf{3.18} &
  \textbf{11.25} &
  \textbf{0.92} \\
 &
  \begin{tabular}[c]{@{}c@{}}Sparse MERIT\\ w/ Expert Balancing Loss\end{tabular} &
  2.23 &
  3.93 &
  3.36 &
  3.12 &
  11.11 &
  \textbf{0.92} \\ \bottomrule
\end{tabular}
}
\label{table:se_load_balance}

\end{table}

Many previous MoE models \cite{shazeer_2017_1, fedus_2022_1}, such as the Switch Transformer, introduce an expert balancing loss to encourage uniform expert utilization. This auxiliary loss penalizes uneven expert usage during training, with the goal of preventing the model from over-relying on a small subset of experts and thus limiting its capacity.

We evaluate the impact of this loss in our MTL setup by conducting an ablation study. Specifically, we adopt the expert balancing loss formulation from the Switch Transformer and compare models trained with and without it. As shown in Table \ref{table:ser_load_balance}, for the SER task, the balancing loss improves performance under the seen noise conditions at -5 dB and 0 dB SNR. However, under most other noise conditions, particularly in unseen environments, models trained without the expert balancing loss achieve better performance.

A similar trend is observed for the SE task, as shown in Table \ref{table:se_load_balance}. While the balancing loss leads to better results under the seen noise condition, it does not improve generalization to unseen noise. We hypothesize that the loss may force experts to be uniformly shared across tasks or input conditions, even when doing so is suboptimal. This could introduce conflicting gradient signals and hinder task-specific specialization, as discussed in \cite{chen_2023_2}. Based on these findings, we do not include the expert balancing loss in our final model.

\section{Conclusions}
\label{sec:conclusion}
This paper proposed Sparse MERIT, a MoE framework for MTL of SE and SER. Sparse MERIT integrates multi-layer self-supervised representations through frame-wise expert routing, enabling task-specific specialization while maintaining a shared backbone. The model uses task-dependent gating networks to select from a shared set of experts at the frame level, improving learning flexibility and reducing interference between tasks.

Experiments on the MSP-Podcast corpus demonstrate that Sparse MERIT significantly improves robustness under noisy conditions. For the SER task, Sparse MERIT achieves a 12.4\% F1-macro improvement over a baseline relying on SE pre-processing and a 3.8\% improvement over a naive MTL baseline at -5 dB SNR on the test data contaminated with Freesound noise, which was not seen during training. Under another unseen noise condition using DNS noise, Sparse MERIT improves F1-macro by 11.6\% compared to the SE pre-processing baseline and by 2.9\% over the naive MTL baseline. In addition to its strong performance under low-SNR and unseen noise, Sparse MERIT also performs competitively under high-SNR scenarios, even on seen noise conditions where the SE pre-processing baseline typically performs better at low SNR. This finding suggests that Sparse MERIT can help prevent the distortion effects that front-end enhancement models may introduce under low-interference conditions. Furthermore, Sparse MERIT consistently outperforms adaptive MTL baselines, including uncertainty-based loss weighting and PCGrad. For the SE task, Sparse MERIT also demonstrates superior performance across commonly used speech quality metrics under the same test conditions. These results indicate that Sparse MERIT architecture improves MTL effectiveness, offering better generalizability and robustness across both tasks.

Despite these improvements, MERIT introduces additional memory overhead during training. Compared to the naive MTL baseline, it requires approximately an additional 5 GB of GPU memory when the batch size is 4. Although the sparse gating mechanism can improve computational efficiency at inference time in multi-GPU settings (e.g., via expert parallelism), the same benefit may be limited when deployed on a single GPU.

In future work, we plan to explore more flexible expert mechanisms, such as allowing the model to dynamically determine the number of experts to activate per frame instead of using a fixed Top-K selection. We also aim to investigate the use of shared experts across tasks and evaluate Sparse MERIT in more complex conditions, including reverberant environments and multilingual speech, to further assess its generalizability. Furthermore, we plan to extend our method to more diverse speech-related multi-task learning scenarios, such as combining speech recognition, speaker identification, or affective attribute prediction, to evaluate its scalability and task-transfer potential.

\section*{Acknowledgments}
{We would like to thank the National Science and Technology Council (NSTC) Taiwan for funding this research.}

% {\appendix[Proof of the Zonklar Equations]
% Use $\backslash${\tt{appendix}} if you have a single appendix:
% Do not use $\backslash${\tt{section}} anymore after $\backslash${\tt{appendix}}, only $\backslash${\tt{section*}}.
% If you have multiple appendixes use $\backslash${\tt{appendices}} then use $\backslash${\tt{section}} to start each appendix.
% You must declare a $\backslash${\tt{section}} before using any $\backslash${\tt{subsection}} or using $\backslash${\tt{label}} ($\backslash${\tt{appendices}} by itself
%  starts a section numbered zero.)}

%{\appendices
%\section*{Proof of the First Zonklar Equation}
%Appendix one text goes here.
% You can choose not to have a title for an appendix if you want by leaving the argument blank
%\section*{Proof of the Second Zonklar Equation}
%Appendix two text goes here.}

\bibliographystyle{IEEEtran}
\bibliography{ref}

\newpage

\section{Biography Section}
% If you have an EPS/PDF photo (graphicx package needed), extra braces are
%  needed around the contents of the optional argument to biography to prevent
%  the LaTeX parser from getting confused when it sees the complicated
%  $\backslash${\tt{includegraphics}} command within an optional argument. (You can create
%  your own custom macro containing the $\backslash${\tt{includegraphics}} command to make things
%  simpler here.)
 
% \vspace{11pt}

% \bf{If you include a photo:}
\vspace{-33pt}
\begin{IEEEbiography}[{\includegraphics[width=1in,height=1.25in,clip,keepaspectratio]{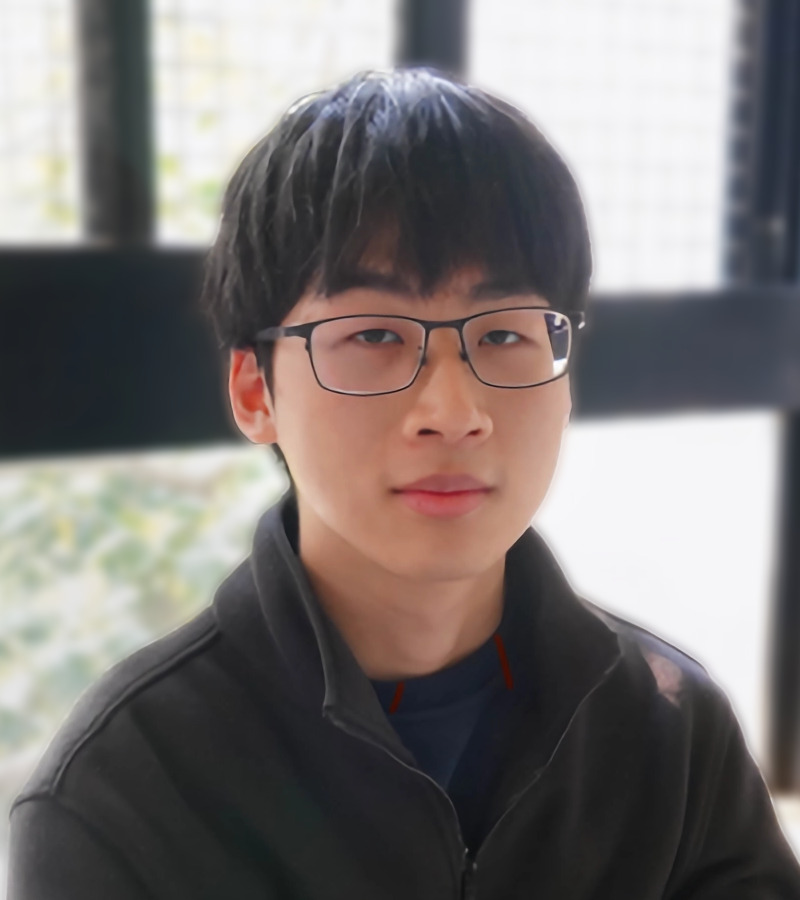}}]{Jing-Tong Tzeng}
(Student Member, IEEE) received the BS degree in Power Mechanical Engineering from National Tsing Hua University, Taiwan, in 2022, and the MS degree from the College of Semiconductor Research, National Tsing Hua University, in 2024. He is currently a research assistant with the Department of Electrical Engineering, National Tsing Hua University. His research interests include speech signal processing, speech emotion recognition, and health analytics. He is a student member of ISCA and the IEEE Signal Processing Society.
\end{IEEEbiography}
\vspace{-20pt}
\begin{IEEEbiography}
[{\includegraphics[width=1in,height=1.25in,clip,keepaspectratio]{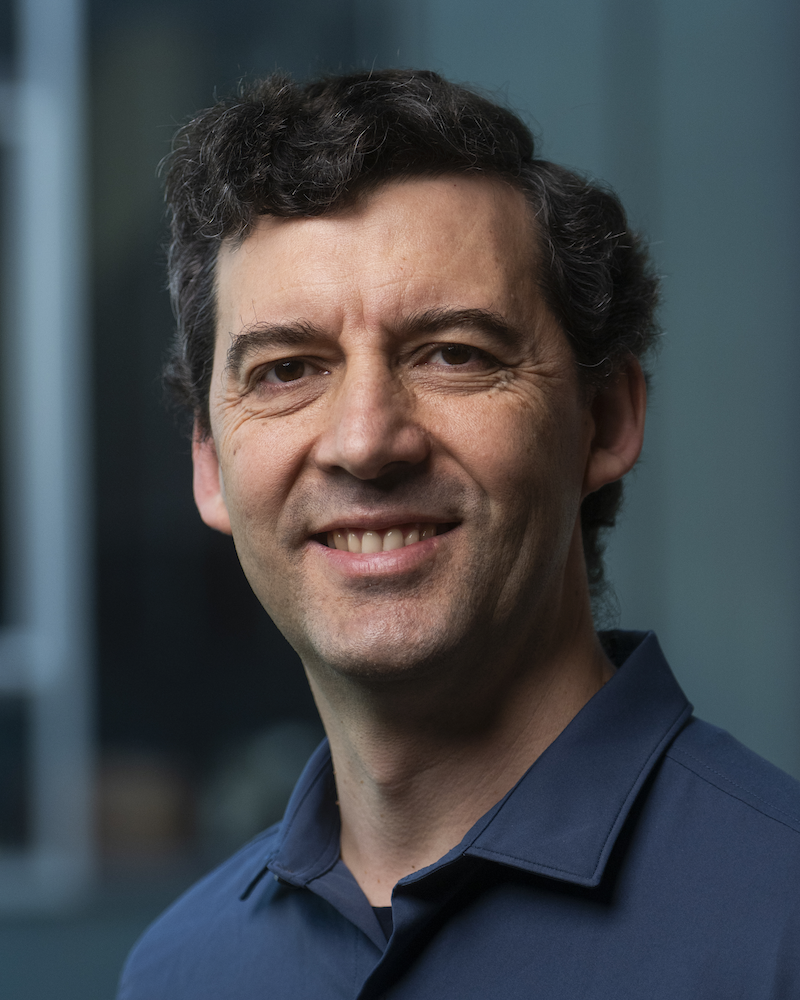}}]{Carlos Busso}
(Fellow, IEEE) is a Professor at Language Technologies Institute, Carnegie Mellon University, where he is also the director of the Multimodal Speech Processing (MSP) Laboratory. He received the BS and MS degrees with high
honors in electrical engineering from the University
of Chile, Santiago, Chile, in 2000 and 2003, respectively, and the PhD degree (2008) in electrical engineering from the University of Southern California (USC), Los Angeles, in 2008. His research interest is in human-centered multimodal machine intelligence and applications, focusing on the broad areas of speech processing, affective computing, multimodal behavior generative models, and foundational models for multimodal processing. He was selected by the School of Engineering of Chile as the best electrical engineer who graduated in 2003 from Chilean universities. He is a recipient of an NSF CAREER Award. In 2014, he received the ICMI Ten-Year Technical Impact Award. His students received
the third prize IEEE ITSS Best Dissertation Award (N. Li) in 2015, and the AAAC Student Dissertation Award (W.-C. Lin) in 2024. He also received the Hewlett Packard Best Paper Award at the IEEE ICME 2011 (with J. Jain), and the Best Paper Award at the AAAC ACII 2017 (with Yannakakis and Cowie). He received the Best of IEEE Transactions on Affective Computing
Paper Collection in 2021 (with R. Lotfian) and the Best Paper Award from IEEE Transactions on Affective Computing in 2022 (with Yannakakis and Cowie). In 2023, he received the Distinguished Alumni Award in the Mid-Career/Academia category by the Signal and Image Processing Institute
(SIPI) at the University of Southern California. He received the 2023 ACM ICMI Community Service Award. He is currently a Senior Area Editor of IEEE/ACM Speech and Language Processing. He is a member of AAAC and a senior member of ACM. He is an IEEE Fellow and an ISCA Fellow.
\end{IEEEbiography}
\vspace{-20pt}
\begin{IEEEbiography}
[{\includegraphics[width=1in,height=1.25in,clip,keepaspectratio]{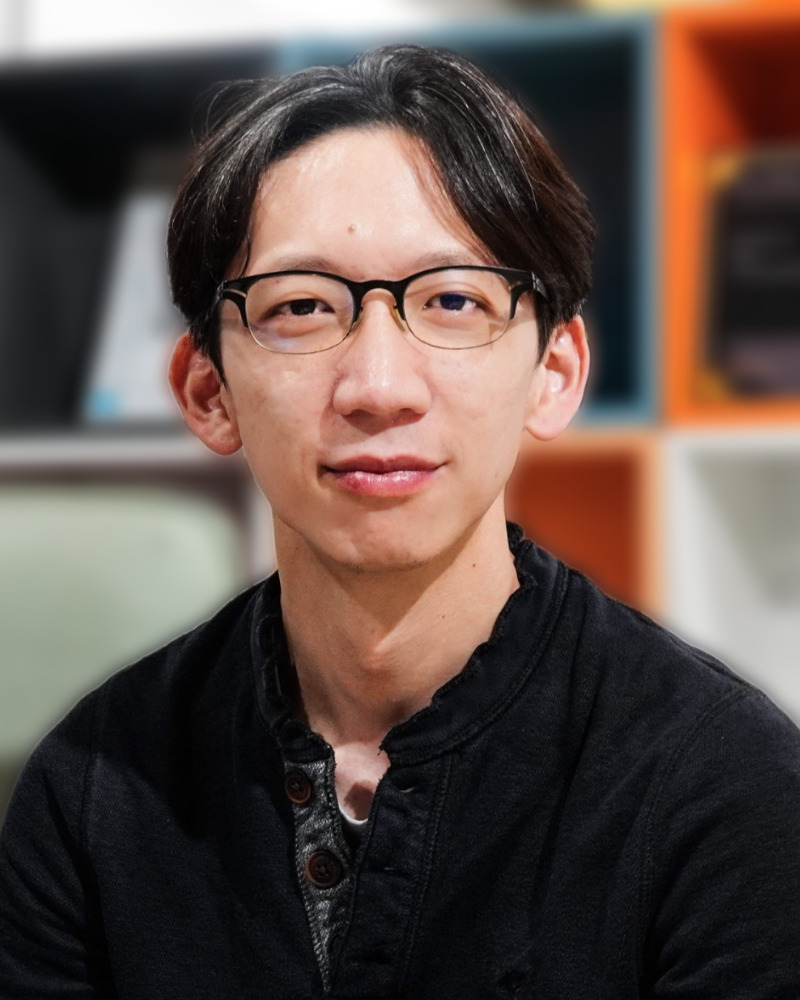}}]{Chi-Chun Lee}
(Senior Member, IEEE) is a Professor at the Department of Electrical Engineering of the National Tsing Hua University (NTHU), Taiwan. He received his BS and PhD degree both in Electrical Engineering from the University of Southern California (USC), USA in 2007 and 2012. His research interests are in speech and language, affective computing, health analytics, and behavioral signal processing. He is an associate editor for the IEEE Transaction on Affective Computing (2020-), the IEEE Transaction on Multimedia (2019-2020), the Journal of Computer Speech and Language (2021-), the APSIPA Transactions on Signal and Information Processing and a TPC member for APSIPA IVM and MLDA committee. He serves as the general chair for ASRU 2023, an area chair for Interspeech 2016, 2018, 2019, senior program committee for ACII 2017, 2019, publicity chair for ACM ICMI 2018, late breaking result chair for ACM ICMI 2023,  sponsorship and special session chair for ISCSLP 2018, 2020. He is the recipient of the the NSTC Outstanding Research Award (2024), the Foundation of Outstanding Scholar’s Young Innovator Award (2020), the CIEE Outstanding Young Electrical Engineer Award (2020), the IICM K. T. Li Young Researcher Award (2020), the NTHU Industry Collaboration Excellence Award (2021), and the MOST Futuretek Breakthrough Award (2018, 2019). He led a team to the 1st place in Emotion Challenge in Interspeech 2009, and with his students won the 1st place in Styrian Dialect and Baby Sound subchallenge in Interspeech 2019. He is a co-author on the best paper award/finalist in Interspeech 2008, Interspeech 2010, IEEE EMBC 2018, Interspeech 2018, IEEE EMBC 2019, APSIPA ASC 2019, IEEE
EMBC 2020, and the most cited paper published in 2013 in Journal of Speech Communication. He is also an ACM and ISCA member.
\end{IEEEbiography}
% \vspace{11pt}

% \bf{If you will not include a photo:}\vspace{-33pt}
% \begin{IEEEbiographynophoto}{John Doe}
% Use $\backslash${\tt{begin\{IEEEbiographynophoto\}}} and the author name as the argument followed by the biography text.
% \end{IEEEbiographynophoto}

\vfill

\end{document}